\documentclass[%
10pt,
amsmath, 
amssymb,
superscriptaddress,
twocolumn,
noshowpacs,
aps,
prl,
longbibliography
]{revtex4-2}

\newcommand{\material}{CeRu\ensuremath{_2}}

\newcommand{\Tc}{T_{\mathrm{c}}}
\newcommand{\HcTwo}{H_{\mathrm{c}2}}
\newcommand{\Hfield}{H}

\newcommand{\kxx}{\kappa_{xx}}
\newcommand{\kxy}{\kappa_{xy}}
\newcommand{\kph}{\kappa_{\mathrm{ph}}}

\newcommand{\kel}{\kappa_{\mathrm{el}}}
\newcommand{\ke}{\kappa_{xx,e}}
\newcommand{\kh}{\kappa_{xx,h}}
\newcommand{\kN}{\kappa_{\mathrm{N}}}

\newcommand{\kb}{k_{\mathrm{B}}}
\newcommand{\Deltae}{\Delta_{e}}
\newcommand{\Deltah}{\Delta_{h}}

\newcommand{\dd}{\mathrm{d}}
\newcommand{\taubd}{\tau_\mathrm{bd}}
\newcommand{\taud}{\tau_\mathrm{d}}
\newcommand{\taueph}{\tau_\mathrm{ph-e}}
\newcommand{\tauephs}{\tau_\mathrm{ph-e,s}}
\newcommand{\tauephn}{\tau_\mathrm{ph-e,n}}

\newcommand{\RRR}{\mathrm{RRR}}
\newcommand{\Hstar}{H^*}

\newcommand{\biasfield}{\SI{50}{\oersted}}
\newcommand{\normalfield}{\SI{9}{\tesla}}
\newcommand{\lowfieldcutoff}{\SI{1}{\tesla}}
\newcommand{\lowesttemperature}{\SI{2.1}{\kelvin}}

\newcommand{\holeweight}{\ensuremath{0.4}}

\newcommand{\twobphpathval}{0.9\pm0.1}
\newcommand{\twobgammaephval}{\ensuremath{200\pm20}}

\newcommand{\singlegap}{\SI{12.1+-0.1}{\kelvin}}
\newcommand{\betaZeroval}{0.46 \pm 0.04}
\newcommand{\betaOneval}{2.9 \pm 0.1}
\newcommand{\Tsval}{5.2 \pm 0.3}
\newcommand{\largegapval}{13.0 \pm 0.1}
\newcommand{\smallgapval}{8.1 \pm 0.2}
\newcommand{\eholeval}{0.83 \pm 0.02}
\newcommand{\hholeval}{0.18 \pm 0.02}
\newcommand{\sommerfeldval}{28.1 \pm 0.3}
\newcommand{\sommerfeldresidueval}{0.75 \pm 0.01}

\newcommand{\sommerfeld}{\SI{\sommerfeldval}{\milli\joule\per\mole\per\kelvin\squared}}

\newcommand{\largegap}{\SI{\largegapval}{\kelvin}}
\newcommand{\smallgap}{\SI{\smallgapval}{\kelvin}}

\usepackage[dvipsnames, svgnames, table]{xcolor}
\usepackage[]{hyperref}
\hypersetup{colorlinks = true, linkcolor = PineGreen, citecolor = blue, filecolor = magenta, urlcolor = blue}

\usepackage{graphicx}
\newcommand{\FigRef}[1]{Fig.~\ref{#1}}
\usepackage[inkscapelatex=false]{svg}
\usepackage[section]{placeins}
\graphicspath{{Figures/}}

\usepackage[separate-uncertainty = true]{siunitx}
\usepackage{physics, braket}
\usepackage{bm}

\AtBeginDocument{\RenewCommandCopy\qty\SI}
\DeclareSIUnit{\oersted}{Oe}
\DeclareSIUnit{\angstrom}{\textup{\AA}}

\usepackage{booktabs}
\usepackage{dcolumn}

\newcommand{\beginsupplement}{
	\clearpage
	\onecolumngrid
	
	\section*{Supplemental Material}
	
	\setcounter{section}{0}
	\renewcommand{\thesection}{S\arabic{section}}%
	\setcounter{table}{0}
	\renewcommand{\thetable}{S\arabic{table}}%
	\setcounter{figure}{0}
	\renewcommand{\thefigure}{S\arabic{figure}}%
	
	\twocolumngrid
}

\begin{document}

\title{%
	\texorpdfstring{%
		Band-Resolved Superconducting Gaps via the Thermal Hall Effect in \material{}}{%
		Band-Resolved Superconducting Gaps via the Thermal Hall Effect in CeRu2}}

\newcommand{\UCB}{Department of Physics, University of California, Berkeley, California 94720, USA}
\newcommand{\LBNL}{Materials Sciences Division, Lawrence Berkeley National Laboratory, Berkeley, California 94720, USA}
\newcommand{\CIFAR}{CIFAR Quantum Materials, Toronto, Ontario M5G 1M1, Canada}
\newcommand{\KAVLI}{Kavli Energy NanoScience Institute, Berkeley, California 94720, USA}

\author{Darian \surname{Hall}}
\email[email: ]{darian\_hall314@berkeley.edu}
\affiliation{\UCB}
\affiliation{\KAVLI}

\author{Luke \surname{Pritchard Cairns}}
\author{Nico \surname{Rowland}}
\affiliation{\UCB}

\author{Yuanqi \surname{Lyu}}
\author{Kohtaro \surname{Yamakawa}}
\affiliation{\UCB}

\author{Josue \surname{Rodriguez}}
\affiliation{\UCB}
\affiliation{\LBNL}

\author{Kirin \surname{Chanteloup}}
\affiliation{\UCB}

\author{Didier \surname{Perrodin}}
\affiliation{\LBNL}

\author{Joel E. \surname{Moore}}
\affiliation{\UCB}
\affiliation{\KAVLI}
\affiliation{\LBNL}

\author{James G. \surname{Analytis}}
\thanks{Corresponding author}
\email[email: ]{analytis@berkeley.edu}
\affiliation{\UCB}
\affiliation{\KAVLI}
\affiliation{\LBNL}
\affiliation{\CIFAR}

\date{\today}

\begin{abstract}
Standard bulk probes of multiband superconductors cannot assign pairing gaps to specific electron or hole Fermi surfaces. Here we show that thermal Hall transport ($\kxy$), constrained by specific heat and normal-state magnetotransport, resolves this ambiguity in the compensated heavy-fermion superconductor \material{}. Fixing the gap magnitudes via specific heat ($\Deltae/\kb\approx\SI{13}{\kelvin}$, $\Deltah/\kb\approx\SI{8}{\kelvin}$), and band mobilities completely determines the normal-state electronic degrees of freedom. A low-field sign reversal in $\kxy$ places the larger gap on the heavy Ce $4f$ electron sheets and the smaller on the hole sheets.
\end{abstract}

\maketitle

Many superconductors host multiple Fermi-surface sheets across which the pairing gap can vary substantially~\cite{Suhl1959,Moskalenko1974,Choi2002_mgb2,Souma2003,Yokoya2001,Ding2008}.
This phenomenon is particularly significant in heavy-fermion superconductors, where Kondo hybridization delocalizes $f$-electrons into heavy Fermi-surface sheets that coexist with lighter conduction bands, obscuring how quasiparticle mass renormalization governs participation in the superconducting condensate~\cite{Steglich1979,Pfleiderer2009,Seyfarth2008,Allan2013,Kittaka2014}.
Establishing how gap magnitudes are distributed among distinct bands is essential for linking a material's electronic structure to its pairing mechanism~\cite{Gallagher2016,Hirschfeld2011,Matsuda2006}.
However, standard bulk thermodynamic probes---such as specific heat, penetration depth, and longitudinal thermal conductivity $\kxx$---measure a sum over all sheets weighted by their respective densities of states or velocities~\cite{boakninHeatConductionVortex2003,Bouquet2001,Prozorov2006,Shakeripour2009}.
Because these probes are insensitive to carrier sign, they struggle to determine which Fermi-surface sheet hosts which gap, nor can they unambiguously distinguish multiband pairing from gap anisotropy on a single sheet~\cite{Johnston2013,Fletcher2007,Zehetmayer2013}.

A bulk probe sensitive to carrier sign can directly overcome this limitation. Whereas the electrical Hall conductivity vanishes in the superconducting state, the thermal Hall conductivity $\kxy$ remains finite in the mixed state and selectively tracks the entropy transport of thermally excited Bogoliubov quasiparticles (BQPs) without a competing phonon background~\cite{onoseLorenzNumberCeCoIn52007,checkelsky2012thermal}. Although the exponential freeze-out of BQPs below $\Tc$ is governed by each band's respective gap scale~\cite{bardeenTheoryThermalConductivity1959,tinkham2004introduction}, gaps of comparable magnitude suppress these excitations across overlapping temperature windows. Developing a robust $\kxy$ methodology to disentangle multiple superconducting gaps and assign them to specific Fermi surface sheets is therefore essential for resolving the underlying pairing mechanisms in correlated and flat-band superconductors.

The cubic Laves-phase superconductor \material{} ($\Tc \approx \SI{6}{\kelvin}$) provides an ideal platform for this test. Band-structure calculations and de Haas--van Alphen measurements establish a compensated metal with electron- and hole-like Fermi-surface sheets of equal total volume~\cite{Hedo1995CeRu2}. Carrier compensation renders the thermal Hall response exceptionally sensitive to modest differences between gaps: because the oppositely signed electron and hole contributions nearly cancel in the normal state, the net sign of $\kxy$ is governed by a slight mobility imbalance. Below $\Tc$, the band with the larger gap freezes out more rapidly; if that band dominates the normal-state Hall response, even a moderate gap hierarchy will trigger a complete sign reversal of $\kxy$ upon cooling. In \material{}, photoemission, muon spin-rotation (\si{\micro SR}) spectroscopy, and thermodynamic studies have long pointed to a nonuniform gap structure~\cite{mielkeiiiLocalSpectroscopicEvidence2022a,kissPhotoemissionSpectroscopicEvidence2005,kittakaVerificationAnisotropicWave2013,gerguriDistinctUniaxialStress2026}, yet whether this reflects single-band anisotropy or distinct gaps on separate sheets has remained unresolved for over two decades. Crucially, because both Ru $4d$ and hybridized itinerant Ce $4f$ states contribute at the Fermi level~\cite{Higuchi1996}---with the electron sheets near $\Gamma$ carrying predominantly $4f$ character and the heaviest effective masses---identifying which carrier type hosts the dominant gap directly illuminates the role of $4f$ electrons in the pairing state.

\begin{figure*}[htp]
  \centering
  \includegraphics[width=\linewidth]{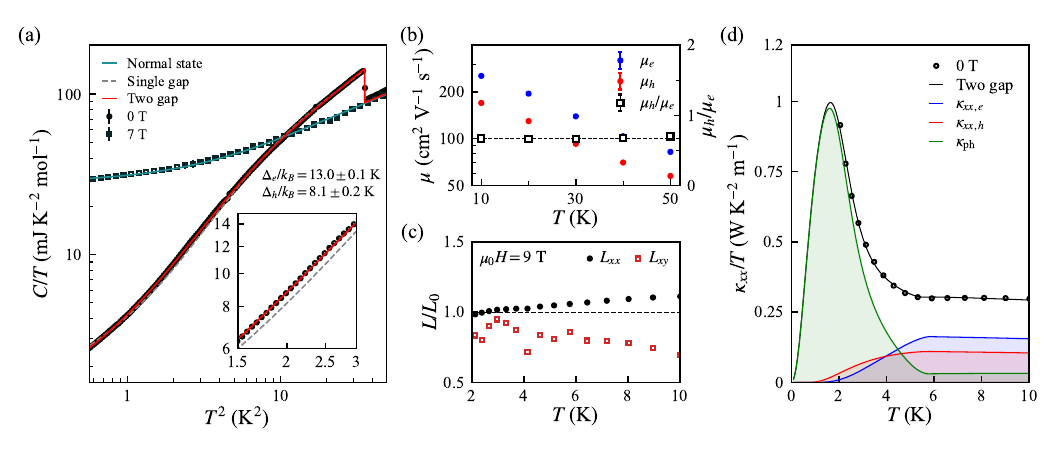}
  \caption{Compensated two-band transport and two-gap superconducting state in \material{}.
  (a) Specific heat $C/T$ in the normal state (squares, teal curve) and superconducting state (circles, red curve), showing fits to a single-gap (gray dashed curve) and two-gap (red curve) BCS model. The inset highlights the low-temperature deviation of the single-gap fit ($\Delta/\kb = \singlegap$).
  (b) Electron and hole mobilities ($\mu_e$ and $\mu_h$) and their ratio extracted from a compensated two-band magnetotransport model, yielding $\mu_h / \mu_e = 0.68$ (dashed line).
  (c) Longitudinal ($L_{xx}$) and transverse ($L_{xy}$) Lorenz ratios at \normalfield{} normalized by the Sommerfeld value $L_0$. The dashed line marks the Wiedemann--Franz expectation ($L/L_0 = 1$).
  (d) Zero-field $\kxx/T$ below $\Tc$, showing the strong enhancement of the phonon thermal conductivity ($\kph$, green) as the electron-band ($\ke$, blue) and hole-band ($\kh$, red) quasiparticle channels are gapped out. The solid black curve is the total thermal conductivity from the combined two-gap BRT and Debye--Callaway model.
  }
  \label{fig:figure_thermo_and_transport}
\end{figure*}

Here we demonstrate that combining thermal Hall transport with zero-field thermodynamics and magnetotransport uniquely assigns superconducting gaps of comparable magnitude to their respective Fermi-surface sheets. In \material{}, zero-field specific heat fixes two distinct temperature-equivalent gap scales ($\SI{8}{\kelvin}$ and $\SI{13}{\kelvin}$, respectively) while the normal-state mobility ratio $\mu_h/\mu_e$ locks in their relative transport weights, completely constraining the two-band Bardeen--Rickayzen--Tewordt (BRT) electronic channels in $\kxx$ against a pronounced phonon enhancement. Extending this parameter-independent BRT decomposition to $\kxy$ via band-dependent thermal Hall angles, we show that a prominent low-field sign reversal of $\kxy$ upon cooling from an electron-dominated normal state unambiguously places the larger gap on the heavy, Ce $4f$-derived electron sheets and the smaller gap on the hole sheets, before giving way at higher fields to a quadratic mixed-state recovery up to the upper critical field $\HcTwo$.

We begin by characterizing the bulk superconducting state of \material{} single crystals [see Supplemental Material (SM) for experimental methods]. Figure~\ref{fig:figure_thermo_and_transport}(a) displays the specific heat in both the superconducting (circles, $\mu_0H=\SI{0}{\tesla}$) and normal (squares, $\mu_0H=\SI{7}{\tesla}$) states. A fit to the normal-state data yields a Sommerfeld coefficient $\gamma_n = \sommerfeld$ ~\cite{Huxley_1993_CeRu2,hedo1998superconducting}, reflecting a moderately enhanced electronic density of states. Subtracting the lattice contribution from the \SI{0}{\tesla} data (teal curve, see SM for more details) isolates the electronic specific heat, which exhibits clear low-temperature deviations from an isotropic single-gap BCS description [\FigRef{fig:figure_thermo_and_transport}(a) and inset, gray dashed curve]. Instead, a two-gap BCS model~\cite{tinkham2004introduction} accurately captures the data across the full temperature range [\FigRef{fig:figure_thermo_and_transport}(a), red curve], yielding a dominant gap $\Delta_e/\kb = \largegap{}$ and a smaller gap $\Delta_h/\kb = \smallgap{}$ (whose respective electron and hole assignments are established by $\kxy$ below). While $\Delta_e$ lies near the upper bound of previously reported gap scales in \material{} ($8.1$--$\SI{12.8}{\kelvin}$ and $2.3$--$\SI{5.8}{\kelvin}$), $\Delta_h$ exceeds earlier estimates~\cite{gerguriDistinctUniaxialStress2026,kissPhotoemissionSpectroscopicEvidence2005,mielkeiiiLocalSpectroscopicEvidence2022a,kittakaVerificationAnisotropicWave2013}. This difference arises because prior studies typically characterized the order parameter using an anisotropic single-band model, and because the modest gap ratio ($\Delta_e/\Delta_h \approx 1.6$) produces subtle thermodynamic inflection points that are easily obscured without high-resolution low-temperature data.

To constrain how these two gaps partition the electronic heat current, we extract the normal-state carrier parameters via magnetotransport (see SM). Fitting the longitudinal and Hall conductivities to a compensated two-band model ($n_e = n_h$) yields a total carrier density $n = \SI{3.6e21}{\per\centi\meter\cubed}$ and a temperature-independent mobility ratio $\mu_h/\mu_e = 0.68$ [\FigRef{fig:figure_thermo_and_transport}(b)]. We next connect the normal-state charge and entropy currents through the Lorenz ratios $L_{ij} = \kappa_{ij}/(\sigma_{ij}T)$ for $ij \in \{xx, xy\}$. Figure~\ref{fig:figure_thermo_and_transport}(c) compares $L_{xx}$ and $L_{xy}$ at \normalfield{} against the Sommerfeld value $L_0 \approx \SI{2.44e-8}{\watt\ohm\per\kelvin\squared}$. On cooling below \SI{10}{\kelvin}, $L_{xx}$ approaches $L_0$ from above while $L_{xy}$ remains comparable in magnitude with a modest suppression at higher temperatures, confirming a predominantly electronic normal-state heat current. Crucially, the low-temperature recovery of the Wiedemann--Franz law ($L_{xx} \to L_0$) determines the total normal-state electronic thermal conductivity $\kel = L_0 \sigma_{xx} T$ directly from the \normalfield{} electrical conductivity, while carrier compensation fixes the relative hole-to-electron normal-state thermal conductance ratio to $\mu_h/\mu_e = 0.68$.

In zero field, $\kxx/T$ exhibits a pronounced enhancement below $\Tc$ [\FigRef{fig:figure_thermo_and_transport}(d)], driven by the interplay between the collective phonon and electronic channels. Because the superconducting condensate carries no entropy, electronic heat transport below $\Tc$ is mediated exclusively by thermally excited Bogoliubov quasiparticles (BQPs). For an $s$-wave superconductor, the BQP entropy content and its resulting thermal conductivity are governed by Bardeen--Rickayzen--Tewordt (BRT) theory~\cite{bardeenTheoryThermalConductivity1959}. Simultaneously, as the opening of the superconducting gaps depletes the BQP population, phonon-electron scattering is strongly suppressed, elongating the phonon mean free path and driving the observed surge in the lattice thermal conductivity $\kph$ [\FigRef{fig:figure_thermo_and_transport}(d), green curve].

Crucially, our normal-state transport and specific heat measurements completely determine the electronic thermal conductivity below $\Tc$ without free parameters. We model the electronic channel as the sum of two independent $s$-wave BRT contributions corresponding to the electron and hole bands, highlighted by the blue ($\ke$) and red ($\kh$) curves in \FigRef{fig:figure_thermo_and_transport}(d). The magnitudes and temperature dependencies of the respective gaps are fixed directly by our zero-field specific heat data, while the fractional hole-band weight (\holeweight) is fixed by the normal-state mobility ratio $\mu_h/\mu_e$ (see SM), reflecting a compensated carrier system with a small, negative normal-state Hall coefficient.

With the electronic channels $\ke$ and $\kh$ fully constrained ($\kel=\ke+\kh$), the remaining degrees of freedom reside entirely within the lattice channel $\kph = \kxx - \kel$. We capture $\kph$ accurately using a Debye--Callaway model that incorporates the BRT-prescribed reduction in phonon-BQP scattering below $\Tc$ [\FigRef{fig:figure_thermo_and_transport}(d), solid black curve; see SM]. Furthermore, attributing the low-temperature thermal conductivity to a two-band BQP channel rather than an unphysically large phonon background is independently corroborated by the rapid recovery of $\kxx(\Hfield)$ in the mixed state below $0.75\HcTwo$, where vortex scattering would otherwise strongly damp lattice transport (see SM). Deconstructing $\kxx$ into its band-resolved electronic ($\ke, \kh$) and phononic ($\kph$) components provides the direct foundation needed to link our longitudinal transport to the thermal Hall conductivity $\kxy$ via the thermal Hall angle.

\begin{figure}
    \centering
    \includegraphics[width=\linewidth]{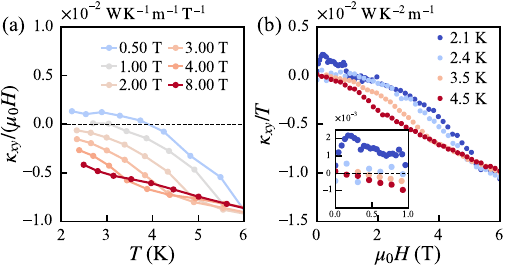}
    \caption{Thermal Hall conductivity of \material{} in the mixed state.
    (a) Temperature dependence of $\kxy/(\mu_0\Hfield)$ showing a prominent sign reversal at \SI{0.5}{\tesla} that is rapidly suppressed with increasing field. Lines are guides to the eye.
    (b) Field dependence of $\kxy / T$, exhibiting a quadratic-in-field recovery toward the normal state. At \SI{2.1}{\kelvin}, a low-field peak emerges below \SI{1}{\tesla} (magnified in the inset).
    } \label{fig:figure_kxy_data}
\end{figure}

\begin{figure*}[t]
    \centering
    \includegraphics[width=0.86\linewidth]{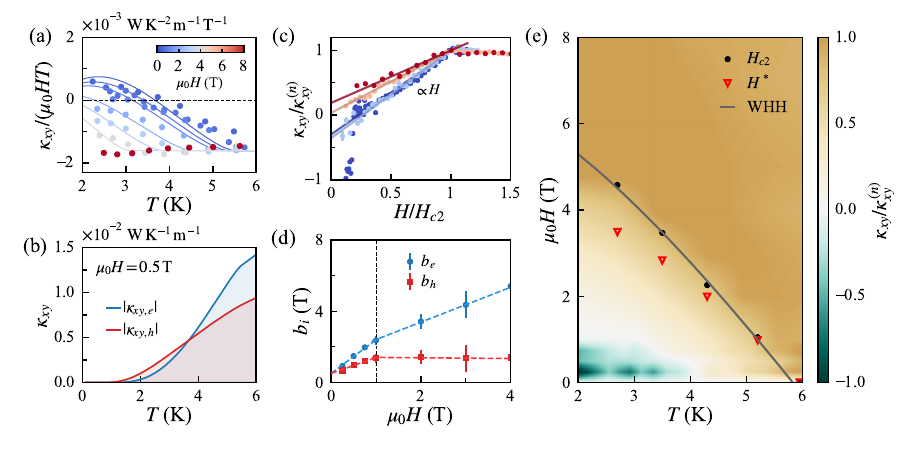}
    \caption{Two-band origin of the mixed-state thermal Hall sign reversal and $\Hfield$--$T$ phase diagram of \material{}.
  (a) Temperature dependence of $\kxy/(\mu_0\Hfield T)$ in the mixed state at fixed fields, tracking the low-field sign reversal below \lowfieldcutoff{}. Solid curves are two-band fits to Eq.~\eqref{eq:two_band_brt_kxy_T}.
  (b) Decomposed electron (blue) and hole (red) contributions to the thermal Hall conductivity at \SI{0.5}{\tesla}.
  (c) Thermal Hall conductivity normalized by the extrapolated normal-state behavior, $\kxy/\kxy^{(n)}$. Solid lines are linear fits to Eq.~\eqref{eq:kxy_H_phenom} in the intermediate-field regime.
  (d) Field dependence of the two-band fitting coefficients $b_e$ and $b_h$; dashed lines are guides to the eye.
  (e) $\Hfield$--$T$ phase diagram of \material{} overlaid on an intensity map of $\kxy/\kxy^{(n)}$.
  Black circles and red triangles mark $\HcTwo$ and the magnetic irreversibility ($\Hstar$) extracted from magnetization (see SM), respectively.
  The light gray curve is a Werthamer--Helfand--Hohenberg (WHH) fit to the upper critical field $\HcTwo$~\cite{Werthamer_Helfand_Hohenberg_1966}.
  } \label{fig:figure_kxy}
\end{figure*}

We next examine $\kxy$, which selectively probes electronic excitations and their signed band contributions without a competing phonon background. In the normal state, $\kxy/(\mu_0\Hfield T) = -\lambda_0$ is field- and temperature-independent~\cite{AshcroftMermin1976}, with $\lambda_0 \approx \SI{1.6e-3}{\watt\per\kelvin\squared\per\meter\per\tesla}$, and the transverse Lorenz ratio remains close to $L_0$ [\FigRef{fig:figure_thermo_and_transport}(c)].
This confirms that $\kxy$ is governed exclusively by charge carriers; moreover, below $\Tc$, the exponential freeze-out of both signed channels as $T\rightarrow0$ [\FigRef{fig:figure_kxy}(b)]---in stark contrast to the large phonon peak in $\kph$ [\FigRef{fig:figure_thermo_and_transport}(d)]---rules out any anomalous phonon skew-scattering contribution in the vortex state.
Entering the superconducting state at low fields, $\kxy$ tracks the thermally excited BQP populations across the two carrier bands, resolving a prominent sign inversion below \lowfieldcutoff{} on cooling [\FigRef{fig:figure_kxy_data}(a)]. With increasing field, this sign reversal is rapidly suppressed and gives way to a quadratic-in-field recovery of $\kxy/T$ toward the normal state [\FigRef{fig:figure_kxy_data}(b)], accompanied at \lowesttemperature{} by a sharp low-field peak below \SI{1}{\tesla} [\FigRef{fig:figure_kxy_data}(b), inset] that mirrors previously measured heavy-fermion superconductors~\cite{onoseLorenzNumberCeCoIn52007,Cairns2021}.

To quantify how the band-resolved longitudinal channels govern this behavior, we link $\kxy$ to our zero-field BRT decomposition via the band-dependent thermal Hall angles $\tan\theta_i(\Hfield) = \kappa_{xy,i} / \kappa_{xx,i}$:
\begin{equation}
  \frac{\kxy}{T} =
      \lambda_0\left[-b_e(\Hfield)R_e(T) + b_h(\Hfield)R_h(T)\right],
  \label{eq:two_band_brt_kxy_T}
\end{equation}
where $R_{e,h}(T) \leq 1$ are the parameter-free BRT suppression factors fixed by specific heat for the electron and hole gaps. Here, the field-dependent weights $b_i(\Hfield) = (\kappa_{xx,i}^{(n)}/\lambda_0 T)\lvert\tan\theta_i(\Hfield)\rvert$ encapsulate the normal-state band conductivities $\kappa_{xx,i}^{(n)}$ (fixed by $\mu_h/\mu_e$) and their respective Hall angles.
The factorization of the field and temperature dependencies in Eq.~\eqref{eq:two_band_brt_kxy_T} is well justified in the low-field limit ($H\ll \HcTwo$). Because our crystals lie in the moderately dirty limit ($\ell_i\leq\xi_0$; see SM), elastic impurity scattering dominates the normal-state relaxation rate, pinning $\tau_i$ below $\Tc$ so that the temperature evolution of $\kxy/T$ at fixed field is governed by the BQP freeze-out $R_i(T)$. Fitting Eq.~\eqref{eq:two_band_brt_kxy_T} over $\SI{2.3}{\kelvin} < T < \SI{5.8}{\kelvin}$ under the normal-state boundary condition $b_e(\Hfield) - b_h(\Hfield) = \mu_0\Hfield$ accurately captures the low-field temperature evolution and sign reversal [\FigRef{fig:figure_kxy}(a,b)]. Remarkably, the extracted coefficients $b_e(\Hfield)$ and $b_h(\Hfield)$ reveal a crossover near \SI{1}{\tesla} from compensated two-carrier transport to an electron-dominated regime [\FigRef{fig:figure_kxy}(d)], directly accounting for the low-field peak observed at base temperature in \FigRef{fig:figure_kxy_data}(b).

At intermediate fields ($\SI{1}{\tesla} < \mu_0\Hfield < \mu_0\HcTwo$), the quadratic-in-field recovery of $\kxy$ seen in \FigRef{fig:figure_kxy_data}(b) translates to a linear variation of the normalized response $\kxy/\kxy^{(n)}$ with field [\FigRef{fig:figure_kxy}(c)]. Because $\kxy^{(n)} = -\lambda_0\mu_0\Hfield T$ is purely linear in $\Hfield$, this intermediate-field regime---reflecting an odd-in-field $\Hfield\lvert\Hfield\rvert$ contribution alongside a residual linear term---is compactly parameterized by a single dimensionless slope $D(T) \geq 0$:
\begin{equation}
    \frac{\kxy}{\kxy^{(n)}} \simeq 1 - D(T)\left(1 - \frac{\lvert\Hfield\rvert}{\HcTwo}\right).
    \label{eq:kxy_H_phenom}
\end{equation}
As shown by the linear fits in \FigRef{fig:figure_kxy}(c), $D(T)$ increases on cooling, driving the extrapolated zero-field intercept $1 - D(T)$ to reverse sign near \SI{3.5}{\kelvin} before saturating at \SI{2.1}{\kelvin}. While this intermediate-field intercept should not be conflated with the true zero-field Hall response, its sign reversal below \SI{3.5}{\kelvin} is a direct signature of the smaller superconducting gap on the hole band: in a single-band superconductor, a linear mixed-state recovery of $\kxy/\Hfield$ should simply vanish as $H, T \to 0$ without changing sign.

Microscopically, because electron and hole bands enter $\kxy$ with opposite signs and scale as $\kappa_{xy,i} \propto \Hfield R_i \tau_i^2$, the observed $\Hfield\lvert\Hfield\rvert$ term can arise from the field dependence of either the BQP population ($R_i$) or the vortex-scattering relaxation time ($\tau_i$). Notably, the simple substitution of a mean-field gap suppression $\Delta_i(\Hfield) \propto \sqrt{1 - \lvert\Hfield\rvert/\HcTwo}$ into the standard BRT integral yields $1 - R_i \propto (1 - \lvert\Hfield\rvert/\HcTwo)^{3/2}$ near $\HcTwo$, which has a vanishing slope at $\HcTwo$ and fails to produce the linear behavior in Eq.~\eqref{eq:kxy_H_phenom}. By contrast, the microscopic mixed-state treatment of Caroli and Cyrot~\cite{Caroli1965} yields a leading-order correction proportional to the spatially averaged order parameter $\langle\lvert\Delta\rvert^2\rangle \propto 1 - \lvert\Hfield\rvert/\HcTwo$, directly matching our experimental field dependence. Establishing how far below $\HcTwo$ this linear regime extends in a multiband vortex lattice remains an important open question for theory.

Figure~\ref{fig:figure_kxy}(e) summarizes our results in an $\Hfield$--$T$ phase diagram of $\kxy/\kxy^{(n)}$, overlaid with $\HcTwo(T)$ and the magnetic irreversibility field $\Hstar(T)$ from \FigRef{fig:figure_mag}. Constrained by two-gap specific heat and normal-state magnetotransport, our thermal Hall measurements establish that the larger superconducting gap in \material{} resides on the electron band, while BQP excitations across the smaller hole gap govern the low-field sign reversal at low temperatures. At intermediate fields, vortex-state transport gives way to a linear Caroli--Cyrot recovery of $\kxy/\kxy^{(n)}$ toward the normal state, where both longitudinal and transverse Lorenz ratios converge to $L_0$ above $\HcTwo$. More broadly, whereas thermodynamic probes and longitudinal heat transport are insensitive to carrier sign---and often obscured by dominant phonon backgrounds---our work unifies parameter-free longitudinal BRT channels with band-dependent thermal Hall angles to isolate signed quasiparticle currents. This approach presents a general blueprint for assigning carrier bands to superconducting gaps and disentangling multiband quasiparticle dynamics across compensated, flat-band, and heavy-fermion superconductors.

\textit{Note added.---}During the completion of this manuscript, we became aware of a related preprint on the thermal Hall effect in elemental Nb~\cite{zhang_2026_thermal_hall_niobium}, which similarly observes a low-field sign reversal of $\kxy$ below $\Tc$ arising from two-gap pairing on electron- and hole-like bands. The two studies are highly complementary: whereas Nb is strongly uncompensated with gaps differing by nearly an order of magnitude extracted from longitudinal fits, \material{} is a compensated system where two gaps of comparable magnitude are fixed independently by specific heat, enabling our explicit two-band BRT decomposition of $\kxy(T, \Hfield)$ across the mixed state.

\subsection*{ACKNOWLEDGMENTS}
This work was supported through a MURI project supported by the Air Force Office of Scientific Research (AFSOR) under grant number FA9550-22-1-0270. Work by Y.L. and L.P.C. was supported by the Brown Institute for Basic Sciences at Caltech.

%\bibliography{aps_bib.bib}
\providecommand{\noopsort}[1]{}\providecommand{\singleletter}[1]{#1}%

\beginsupplement

\subsection{Synthesis} \label{sec:methods}
Single crystals of \material{} were grown by the Czochralski method in a tetra-arc furnace equipped with a water-cooled copper hearth under a high-purity argon atmosphere. A \SI{3.7}{\gram} stoichiometric charge of elemental Ce and Ru ($1{:}2$ molar ratio) was melted and flipped 11 times prior to growth to ensure compositional homogeneity. The single crystal boule was pulled at \SI{4.2}{\milli\meter\per\hour} with a clockwise hearth rotation of \SI{9.8}{rpm} and a counterclockwise seed rotation of \SI{2.2}{rpm}.

Samples for magnetometry and transport were sectioned from the as-grown boule using a wire saw. Across crystals from the growth, the residual resistivity ratio $\RRR = \rho_{xx}(\SI{300}{\kelvin}) / \rho_0$---where $\rho_0$ is the low-temperature normal-state residual resistivity extrapolated to \SI{0}{\kelvin} from above $\Tc$---ranges from 5--8. Bulk magnetization and susceptibility measurements were first performed on a cylindrical disk ($\Tc=\SI{5.8}{\kelvin}$) of diameter \SI{2}{\milli\meter} and thickness \SI{0.44}{\milli\meter}. To eliminate sample-to-sample variations between bulk and transport probes, this same disk was subsequently sectioned into a rectangular bar of dimensions $\SI{2}{\milli\meter} \times \SI{1}{\milli\meter} \times \SI{0.15}{\milli\meter}$ for electrical and thermal transport measurements.

\subsection{Transport and magnetometry measurements}
Bulk magnetization measurements were carried out in a Quantum Design Magnetic Property Measurement System (MPMS) operated in vibrating sample magnetometer (VSM) mode, while electrical and thermal transport measurements were performed in a Quantum Design Physical Property Measurement System (PPMS). Thermal conductivity and thermal Hall measurements were conducted using a home-built steady-state one-heater, three-thermometer probe. A longitudinal heat current $J_Q$ was generated via a resistive chip heater driven by a Stanford Research Systems CS580 voltage-controlled current source, and the resulting longitudinal ($\Delta T_x$) and transverse ($\Delta T_y$) temperature gradients were measured using calibrated Lakeshore Cernox 1050 resistance temperature detectors (RTDs). Each thermometer was excited with a \SI{100}{\nano\ampere} AC current at \SI{133.77}{\hertz} sourced by a Keithley 6221, and the differential voltages were read out using Zurich Instruments MFLI lock-in amplifiers.

To account for geometric contact misalignment, the measured longitudinal and transverse thermal (electrical) resistivities were field-symmetrized and antisymmetrized, respectively, using positive and negative magnetic field sweeps. The thermal (electrical) conductivity tensor components were then obtained by matrix inversion.
To ensure a uniform vortex state and avoid flux-trapping history effects between field polarities, the sample was thermally cycled above $\Tc$ and zero-field cooled (ZFC) prior to acquiring each positive and negative field sweep. Similarly, fixed-field temperature sweeps were recorded under ZFC conditions by cooling the sample in zero field to base temperature before applying the target magnetic field and warming through $\Tc$.

\subsection{Compensated two-band magnetotransport}
\begin{figure}[htp!]
    \centering
    \includegraphics[width=\linewidth]{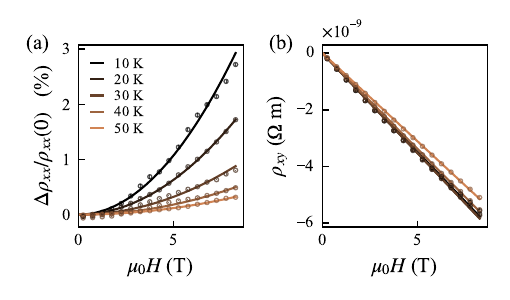}
    \caption{Normal-state magnetotransport of \material{} from $\SI{10}{\kelvin}$ to $\SI{50}{\kelvin}$. (a) Longitudinal magnetoresistance $\Delta\rho_{xx}/\rho_{xx}(0)$ and (b) Hall resistivity $\rho_{xy}(\Hfield)$ (circles) alongside simultaneous two-band fits (solid curves) to Eqs.~\eqref{eq:sm_sigma_xx}--\eqref{eq:sm_sigma_xy}.
    }\label{fig:magnetotransport}
\end{figure}

To determine the normal-state carrier density and band mobilities of \material{}, we simultaneously fit the field-dependent magnetoresistance $\Delta\rho_{xx}/\rho_{xx}(0)$ and Hall resistivity $\rho_{xy}(\Hfield)$ across multiple temperatures from $\SI{10}{\kelvin}$ to $\SI{50}{\kelvin}$ [\FigRef{fig:magnetotransport}(a,b)]. Motivated by band-structure calculations and de Haas--van Alphen measurements establishing \material{} as a compensated metal~\cite{Hedo1995CeRu2,Higuchi1996}, we enforce exact carrier compensation ($n_e = n_h = n_\mathrm{tot}/2$). Within the semiclassical two-band Drude model, the longitudinal and transverse electrical conductivities are given by
\begin{align}
    \sigma_{xx}(B) &= \frac{n_\mathrm{tot} e}{2} \left( \frac{\mu_e}{1 + (\mu_e B)^2} + \frac{\mu_h}{1 + (\mu_h B)^2} \right), \label{eq:sm_sigma_xx} \\
    \sigma_{xy}(B) &= \frac{n_\mathrm{tot} e B}{2} \left( -\frac{\mu_e^2}{1 + (\mu_e B)^2} + \frac{\mu_h^2}{1 + (\mu_h B)^2} \right), \label{eq:sm_sigma_xy}
\end{align}
where $B = \mu_0 \Hfield$, $e$ is the elementary charge, and $\mu_e(T)$ and $\mu_h(T)$ are the electron and hole mobilities, respectively. Here we use the sign convention $\rho_{xy}(B) = \sigma_{xy}/(\sigma_{xx}^2 + \sigma_{xy}^2)$ and $\rho_{xx}(B) = \sigma_{xx}/(\sigma_{xx}^2 + \sigma_{xy}^2)$, such that both $\sigma_{xy}$ and $\rho_{xy}$ are negative for electrons and positive for holes.

At $\SI{10}{\kelvin}$, we obtain $n_e = n_h = \SI{1.8e21}{\per\centi\meter\cubed}$ ($n_\mathrm{tot} = \SI{3.6e21}{\per\centi\meter\cubed}$) and mobilities $\mu_e = \SI{254}{\centi\meter\squared\per\volt\per\second}$ and $\mu_h = \SI{170}{\centi\meter\squared\per\volt\per\second}$. Because $\mu_e > \mu_h$, the electron band dominates the normal-state Hall response, accounting for the negative slope of $\rho_{xy}(\Hfield)$ in \FigRef{fig:magnetotransport}(b). Below \SI{50}{\kelvin}, the mobility ratio remains nearly constant at
\begin{equation}
    r \equiv \frac{\mu_h}{\mu_e} \approx 0.68.
\end{equation}

This mobility ratio directly constrains several normal-state and superconducting transport parameters:

\paragraph{1. Zero-field electronic thermal conductivity weights.}
In the low-temperature normal state ($B \to 0$), where the Wiedemann--Franz law $\kappa_{xx,i}^{(n)} = L_0 \sigma_{xx,i}^{(n)} T$ holds for each band, $i \in \{e, h\}$, carrier compensation ($n_e = n_h$) fixes the ratio of the hole- to electron-band normal-state thermal conductivities solely through the mobility ratio:
\begin{equation}
    \frac{\kappa_{xx,h}^{(n)}}{\kappa_{xx,e}^{(n)}} = \frac{n_h e \mu_h}{n_e e \mu_e} = \frac{\mu_h}{\mu_e} = r.
\end{equation}
Consequently, one finds the fractional hole weight $c_h = \sigma_{xx,h}^{(n)} / (\sigma_{xx,e}^{(n)} + \sigma_{xx,h}^{(n)})$ in the zero-field BRT decomposition is fixed by
\begin{equation}
    c_h = \frac{r}{1 + r} \approx \holeweight, \qquad c_e = 1 - c_h \approx 0.60,
\end{equation}
in remarkable agreement with the value obtained when leaving $c_h$ as an unconstrained fitting parameter in $\kxx(T)$ below $\Tc$.

\paragraph{2. Effective mass and quasiparticle scattering time ratios.}
Combining the magnetotransport mobility ratio with our specific heat analysis yields the relative effective masses and relaxation times of the two carrier types. Within a spherical Fermi-surface approximation, the volumetric Sommerfeld coefficient of band $i$ is $\gamma_{V,i} = (k_\mathrm{B}^2 k_{F,i} / 3\hbar^2) m_i^*$, where $k_{F,i} = (3\pi^2 n_i)^{1/3}$. Because $n_e = n_h$ fixes $k_{F,e} = k_{F,h}$, the ratio of the band effective masses is determined directly by the ratio of the Sommerfeld weights $a_h = \num{\hholeval}$ and $a_e = \num{\eholeval}$ from Table~\ref{tab:hc_table}:
\begin{equation}
    \frac{m_h^*}{m_e^*} = \frac{a_h}{a_e} \approx 0.22.
\end{equation}
This confirms that the electron sheets (which carry predominantly Ce $4f$ character near $\Gamma$) are nearly five times heavier than the hole sheets in an effective two-band model. Using $\mu_i = e\tau_i / m_i^*$, the ratio of the quasiparticle transport relaxation times is
\begin{equation}
    \frac{\tau_h}{\tau_e} = \frac{\mu_h m_h^*}{\mu_e m_e^*} = r \frac{a_h}{a_e} \approx 0.14,
\end{equation}
indicating that holes scatter approximately seven times more strongly than electrons in the normal state. Furthermore these transport parameters establish the microscopic length scales governing mixed-state transport. From the zero-temperature upper critical field $\mu_0H=\SI{6.5}{\tesla}$ [Eq.~\eqref{eq:sm_whh}], the Ginzburg-Landau coherence length is $\xi_0=\sqrt{\Phi_0/[2\pi\mu_0\HcTwo(0)]}\approx\SI{7.1}{\nano\meter}$. Using the carrier density $n_e=n_h=\SI{1.8e21}{\per\centi\meter\cubed}$ ($k_F\approx \SI{3.8}{\nano\meter}$) and the \SI{10}{\kelvin} mobilities, the normal-state electronic mean free paths $\ell_i=\hbar k_F\mu_i/e$ are $\ell_e\approx\SI{6.3}{\nano\meter}$ and $\ell_h\approx\SI{4.2}{\nano\meter}$. Both carrier sheets therefore operate in the moderately dirty limit ($\ell_i\leq \xi_0$), consistent with the residual resistivity ratio $\RRR=\text{5--8}$. This regime directly underpins two key aspects of our analysis: (i) elastic impurity scattering dominates over inelastic scattering below $\Tc$, pinning the baseline relaxation times $\tau_i$ so that the low-field temperature dependence of $\kxy$ in Eq.~\eqref{eq:two_band_brt_kxy_T} is controlled by the zero-field BRT population factors $R_i(T)$ and (ii) the dirty-limit condition $\ell_i\leq\xi_0$ rigorously justifies applying the microscopic Caroli-Cyrot vortex-lattice expansion \cite{Caroli1965} to describe the linear-in-$H$ recovery of $\kxy / \kxy^{(n)}$ in Eq.~\eqref{eq:kxy_H_phenom}.

\paragraph{3. Consistency between $\sigma_{xx}$, $R_H$, and the thermal Hall prefactor $\lambda_0$.}
From the extracted $\SI{10}{\kelvin}$ parameters ($n_e = \SI{1.8e21}{\per\centi\meter\cubed}$, $\mu_e + \mu_h = \SI{424}{\centi\meter\squared\per\volt\per\second}$), the zero-field electrical conductivity is determined by $\sigma_{xx}(0) = n_e e (\mu_e + \mu_h) \approx \SI{1.2e7}{\siemens\per\meter}$, which via the Wiedemann--Franz law predicts a normal-state electronic thermal conductivity
\[
\kel/T = L_0 \sigma_{xx}(0) \approx \SI{0.3}{\watt\per\kelvin\squared\per\meter},
\]
in quantitative agreement with our measured normal-state $\kel/T$.

Similarly, in the low-field limit, the normal-state thermal Hall angle of band $i$ is $\tan\theta_i = (L_{xy}/L_0)\mu_i B$, where the factor $L_{xy}/L_0 \approx 0.7$--$0.85$ [\FigRef{fig:figure_thermo_and_transport}(c)] accounts for the modest suppression of the transverse Lorenz ratio relative to $L_0$. Substituting $b_i(B) = (c_i \kel / \lambda_0 T)\lvert\tan\theta_i\rvert$ into the normal-state boundary condition $b_e(B) - b_h(B) = B$ [with $c_e\mu_e - c_h\mu_h = (\mu_e^2 - \mu_h^2)/(\mu_e + \mu_h) = \mu_e - \mu_h$ for a compensated metal] yields an explicit relation for the normal-state thermal Hall prefactor $\lambda_0 = -\kxy^{(n)}/(B T)$:
\begin{equation}
    \lambda_0 = \frac{L_{xy}}{L_0}\frac{\kel}{T}(\mu_e - \mu_h).
    \label{eq:sm_lambda0_relation}
\end{equation}
Evaluating Eq.~\eqref{eq:sm_lambda0_relation} using our independently measured transport parameters---$\mu_e - \mu_h = \SI{84}{\centi\meter\squared\per\volt\per\second}$, $\kel/T \approx \SI{0.3}{\watt\per\kelvin\squared\per\meter}$, and $L_{xy}/L_0 \approx 0.7$---predicts $\lambda_0 \approx \SI{1.7e-3}{\watt\per\kelvin\squared\per\meter\per\tesla}$, in excellent agreement with the value $\lambda_0 \approx \SI{1.6e-3}{\watt\per\kelvin\squared\per\meter\per\tesla}$ measured directly from $\kxy^{(n)}$. Finally, we note that evaluating the Hall coefficient directly from \FigRef{fig:kappa_WF_comparison} at \SI{6}{\kelvin} yields $R_H\approx-\SI{6e-10}{\meter\cubed\per\coulomb}$; while the thermal analog
\begin{equation}
    R_H = \frac{1}{n_ee}\frac{\mu_h-\mu_e}{\mu_h+\mu_e} = -\frac{\lambda_0}{\sigma_{xx}(0)}\left[\frac{L_{xy}}{L_0}\frac{\kel}{T}\right]^{-1},
\end{equation}
yields $R_H\approx-\SI{5e-10}{\meter\cubed\per\coulomb}$, within experimental agreement.

\subsection{Magnetometry and upper critical field}

\begin{figure}[htp!]
    \centering
    \includegraphics[width=\linewidth]{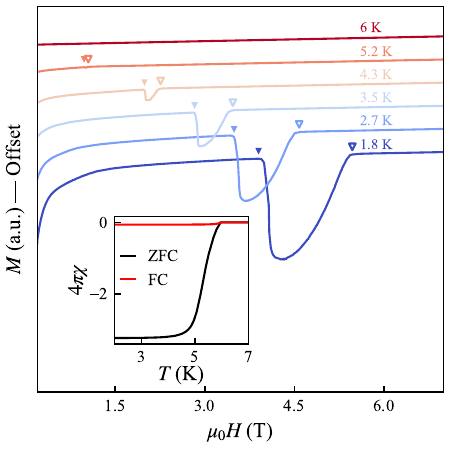}
    \caption{Magnetization curves measured in the superconducting state, showing the magnetic irreversibility field \( \Hstar \) (filled triangles) below the upper critical field \( \HcTwo \) (open triangles). Inset shows zero-field-cooled (ZFC) and field-cooled (FC) magnetic susceptibility measured with a \biasfield{} bias field.
    }\label{fig:figure_mag}
\end{figure}

To characterize the bulk superconducting transition and mixed-state phase boundaries of \material{}, we performed magnetometry measurements (\FigRef{fig:figure_mag}). The inset of \FigRef{fig:figure_mag} displays the zero-field-cooled (ZFC) and field-cooled (FC) magnetic susceptibility $4\pi\chi(T)$, whose sharp diamagnetic onset confirms bulk superconductivity. The ZFC susceptibility $4\pi\chi$ exceeds the ideal diamagnetic value of $-1$ due to the demagnetization factor of the thin, plate-like crystal geometry when the magnetic field is applied parallel to the short cylindrical axis.

Isothermal magnetization curves $M(\Hfield)$ in the main panel of \FigRef{fig:figure_mag} allow us to track both the magnetic irreversibility field $\Hstar(T)$ (filled triangles) and the upper critical field $\HcTwo(T)$ (open triangles).
The field window between $\Hstar(T)$ and $\HcTwo(T)$ widens substantially upon cooling, delineating the pinned vortex regime plotted in the phase diagram of \FigRef{fig:figure_kxy}(e) in the main text.

We model the temperature dependence of the upper critical field using the Werthamer--Helfand--Hohenberg (WHH) theory in the absence of spin-paramagnetic and spin-orbit effects~\cite{Werthamer_Helfand_Hohenberg_1966}:
\begin{equation}
  \ln\left(\frac{T}{\Tc}\right) = \Psi\left(\frac{1}{2}\right) - \Psi\left(\frac{1}{2} + \frac{2\HcTwo(T)}{\pi^2 T \left\lvert \dd\HcTwo/\dd T \right\rvert_{\Tc}}\right),
  \label{eq:sm_whh}
\end{equation}
where $\Psi$ is the digamma function. Fitting the empirically extracted $\HcTwo(T)$ values to Eq.~\eqref{eq:sm_whh} yields an initial slope of $\mu_0 (\dd\HcTwo/\dd T)\rvert_{\Tc} = \SI{-1.62 \pm 0.03}{\tesla\per\kelvin}$, corresponding to a zero-temperature upper critical field $\mu_0\HcTwo(0) = \SI{6.5}{\tesla}$ [light gray curve in \FigRef{fig:figure_kxy}(e)].

\subsection{Specific heat analysis}
Specific heat measurements were performed on a second single crystal with a nearly identical transition temperature \( \Tc = \SI{5.95}{\kelvin} \). In the normal state (measured at $\SI{7}{\tesla}>\mu_0\HcTwo$), the lattice contribution deviates from a simple $T^3$ power law below \( \Tc \) due to phonon dispersion softening~\cite{Huxley_1993_CeRu2,hedo1998superconducting}.
We therefore model the normal-state specific heat above $\HcTwo$ as
\begin{equation}\label{eq:heatcap_normal_state}
    C_n = \gamma_n T + \beta_\mathrm{eff}(T)T^3,
\end{equation}
where $\gamma_n$ is the Sommerfeld coefficient and \( \beta_\mathrm{eff}(T) =\beta_0+\beta_1/[1+(T/T_s)^2] \) is an empirical parametrization of the temperature-dependent Debye coefficient. This corresponds to an effective Debye temperature \( \Theta_D(T)=(12\pi^4 N_\mathrm{at} R/5\beta_\mathrm{eff})^{1/3} \), where \( N_\mathrm{at} = 3 \) is the number of atoms per formula unit and \( R = \SI{8.3145}{\joule\per\mole\per\kelvin} \) is the ideal gas constant. In the zero-temperature limit, $\beta_\mathrm{eff}(0) = \beta_0+\beta_1$ yields a Debye temperature in good agreement with values independently derived from ultrasonic elastic constants~\cite{Wolf1997}.

\begin{table}[htbp]
\centering
\renewcommand{\arraystretch}{1.2}
\begin{tabular}{l c l}
\toprule
Parameter & Value &  \\
\midrule
\(\gamma_n\)  & \num{\sommerfeldval} & \si{\milli\joule\per\mole\per\kelvin\squared}\\
\(\beta_0\)   & \num{\betaZeroval}  & \si{\milli\joule\per\mole\per\kelvin\tothe{4}} \\
\(\beta_1\)   & \num{\betaOneval}   & \si{\milli\joule\per\mole\per\kelvin\tothe{4}} \\
\(T_s\)       & \num{\Tsval}        & \si{\kelvin} \\
\midrule\midrule
\(\Delta_e/\kb\)  & \num{\largegapval}  & \si{\kelvin} \\
\(\Delta_h/\kb\)  & \num{\smallgapval}  & \si{\kelvin} \\
\(a_e\)       & \num{\eholeval}     &  \\
\(a_h\)       & \num{\hholeval}     &  \\
\(\gamma_\mathrm{res}\) & \num{\sommerfeldresidueval} & \si{\milli\joule\per\mole\per\kelvin\squared}\\
\bottomrule
\end{tabular}
\caption{Summary of best-fit parameters and uncertainties for the specific heat of \material{}. The double horizontal line separates parameters obtained from the normal state (\SI{7}{\tesla}, top) from those obtained in the zero-field superconducting state (bottom).}\label{tab:hc_table}
\end{table}

To isolate the electronic specific heat in the zero-field superconducting state, $C_\mathrm{es}(T) = C(T) - \beta_\mathrm{eff}(T)T^3$, we subtract the phonon background $\beta_\mathrm{eff}(T)T^3$ determined from the \SI{7}{\tesla} data. Within BCS theory, the electronic specific heat of a single isotropic gap $\Delta(T)$ is given by~\cite{tinkham2004introduction}
\begin{equation}
\begin{split}
    C_{s}(T, \Delta) = \frac{3\gamma_n}{2\pi^2\kb^3 T^2} \int_0^\infty \left(E^2 - \frac{T}{2}\frac{\dd\Delta^2}{\dd T}\right)\\ \times\frac{\dd\xi}{\cosh^2\!\left[E / (2\kb T)\right]},
\end{split}
\end{equation}
where $E = \sqrt{\xi^2+\Delta^2(T)}$ is the Bogoliubov quasiparticle excitation energy. We approximate the temperature dependence of each gap using the standard BCS interpolation formula $\Delta_i(T)=\Delta_i \tanh\left(\eta\sqrt{\Tc/T-1}\right)$ with $\eta=1.74$~\cite{gross_1986_gap_interpolation}.

To account for multiband pairing across the electron and hole sheets within the well-known $\alpha$-model type heat capacity, we express the total superconducting electronic specific heat as a weighted sum of two independent BCS channels plus a small residual linear contribution $\gamma_\mathrm{res}T$:
\begin{equation}
    C_\mathrm{es} = a_eC_s(T, \Delta_e(T)) + a_hC_s(T, \Delta_h(T)) + \gamma_\mathrm{res}T,
\end{equation}
where $a_e$ and $a_h$ represent the fractional contributions of the electron and hole bands to the normal-state Sommerfeld coefficient $\gamma_n$ (proportional to their respective partial densities of states at the Fermi level). The resulting best-fit parameters and uncertainties are summarized in Table~\ref{tab:hc_table}. Crucially, even without enforcing a strict sum-rule constraint during optimization, the extracted weights satisfy $(a_e + a_h)\gamma_n+\gamma_\mathrm{res} \approx \gamma_n$, confirming entropy conservation and full thermodynamic consistency with the normal-state Sommerfeld coefficient extracted at \SI{7}{\tesla}.

\begin{figure}[htp!]
    \centering
    \includegraphics[width=\linewidth]{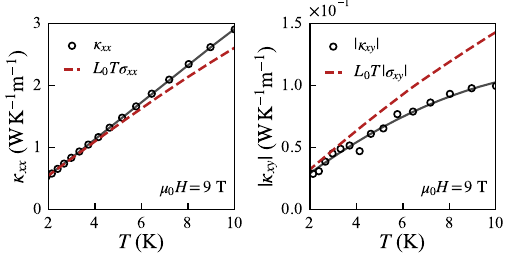}
    \caption{Comparison of the measured thermal conductivities and the normal-state Wiedemann--Franz expectations $L_0 T \sigma_{ij}(T)$ for the longitudinal ($\kxx$, left) and transverse ($\kxy$, right) channels at \normalfield{}. Solid black curves are guides to the eye.
    }\label{fig:kappa_WF_comparison}
\end{figure}

\subsection{BRT model in the mixed state}
With the theory developed by Bardeen, Rickayzen, and Tewordt (BRT)~\cite{bardeenTheoryThermalConductivity1959}, the suppression of the electronic thermal conductivity in an $s$-wave superconductor relative to its normal-state value is governed by the universal function
\begin{equation}
    R(y) = \frac{
    2F_1(-y) + 2y\ln(1+e^{-y}) + y^2/(1+e^y)
    }{
     2F_1(0)
    },
    \label{eq:universal_brt_fn}
\end{equation}
where $y = \Delta(T,\Hfield) / (\kb T)$ is the reduced gap and $F_1(-y)$ is the Fermi--Dirac integral
\begin{equation}
    F_1(-y) = \int_0^\infty \frac{z\,\dd z}{1 + e^{z+y}} = \frac{\pi^2}{12} - \int_0^{y} \ln(1+e^{-z})\,\dd z,
\end{equation}
with $F_1(0) = \pi^2/12$. Neglecting interband scattering between the electron and hole sheets, the total electronic thermal conductivity in the superconducting state is given by the weighted sum of the two bands:
\begin{equation}
    \frac{\ke(T) + \kh(T)}{\kel(T)} = (1 - c_h) R_e(T) + c_h R_h(T),
    \label{eq:sm_two_band_brt}
\end{equation}
where $R_i(T) \equiv R\left(\Delta_i(T)/\kb T\right)$ for $i\in \{e,h\}$. In the zero-field analysis presented in the main text [\FigRef{fig:figure_thermo_and_transport}(d)], the gap magnitudes $\Delta_{e,h}(T)$ are fixed by zero-field specific heat, the fractional hole weight $c_h$ is fixed by the normal-state mobility ratio $\mu_h/\mu_e$, and the total normal-state electronic thermal conductivity $\kel(T) = L_0T\sigma_{xx}(T)$ is obtained from the electrical conductivity measured at \normalfield{} via the Wiedemann--Franz law (\FigRef{fig:kappa_WF_comparison}). Notably, if $c_h$ is instead left as an unconstrained fitting parameter in Eq.~\eqref{eq:sm_two_band_brt}, fitting the zero-field thermal conductivity independently converges to the same hole weight, confirming the self-consistency of our parameter-free decomposition.

The field dependence of $\kxx(\Hfield)$ in the mixed state provides an independent test of this two-band electronic decomposition and rules out attributing the low-temperature thermal conductivity to an unphysically large phonon background. In the mixed state, phonon-vortex scattering reduces the lattice thermal conductivity according to $\kph(\Hfield)=\kph(0)/(1+z\Hfield)$, where $z$ parametrizes the vortex-scattering cross section~\cite{checkelsky2012thermal}. This expression serves as a strict upper bound on $\kph(\Hfield)$ because, in addition to vortex core scattering, the field-induced recovery of the BQP population restores phonon-electron scattering toward its normal-state rate, further damping $\kph(\Hfield)$.

\begin{figure}[htp!]
    \centering
    \includegraphics[width=\linewidth]{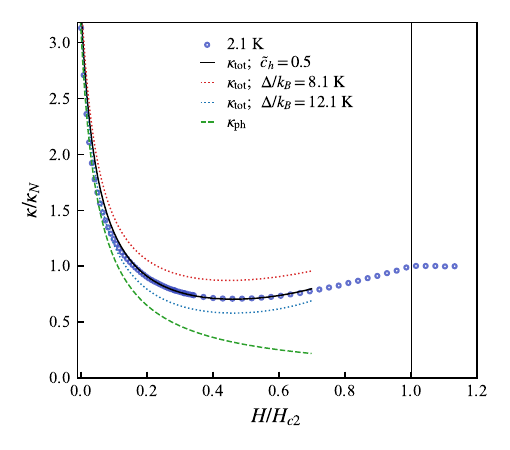}
    \caption{Normalized field dependence of $\kxx(\Hfield)/\kN$ versus $\Hfield/\HcTwo$ at \SI{2.1}{\kelvin}. The initial low-field drop is governed by phonon-vortex scattering (dashed green curve). Dotted blue and red curves show the mixed-state behavior predicted by Eq.~\eqref{eq:sm_kxx_mixed} for a single electronic gap with $\Delta/\kb = \SI{12.1}{\kelvin}$ ($\tilde{c}_h = 0$) and $\Delta/\kb = \SI{8.1}{\kelvin}$ ($\tilde{c}_h = 1$), respectively. The solid black curve shows the two-gap model with $\tilde{c}_h = 0.5$ and gaps specified in Table~\ref{tab:hc_table}, capturing the rapid recovery of the heat current across the mixed state.
    }\label{fig:kxx_hole_bound}
\end{figure}

To model the field evolution of the electronic channels for $\Hfield < 0.75\HcTwo$, we approximate the field-induced suppression of each gap using the Ginzburg--Landau scaling $\Delta_i(\Hfield, T) = \Delta_i(T)\sqrt{1-\Hfield/\HcTwo}$, where $\HcTwo$ is determined from our WHH fit [Eq.~\eqref{eq:sm_whh}]. The total mixed-state longitudinal thermal conductivity is then approximated by
\begin{equation}
    \kxx(\Hfield) \approx \frac{\kph(0)}{1 + z\Hfield} + \kN \sum_{i \in \{e,h\}} \tilde{c}_i R\left(\frac{\Delta_i(\Hfield, T)}{\kb T}\right),
    \label{eq:sm_kxx_mixed}
\end{equation}
where $\kN$ is the normal-state thermal conductivity above $\HcTwo$ and $\tilde c_e + \tilde c_h=1$.

As shown in \FigRef{fig:kxx_hole_bound} at $T=\SI{2.1}{\kelvin}$, attributing the electronic heat current solely to the larger gap ($\tilde c_h=0$, blue dotted curve) severely underestimates the recovery of $\kxx(\Hfield)$ at intermediate fields because the phonon channel (green dashed curve) is rapidly quenched by vortex scattering. Conversely, a single small gap ($\tilde c_h=1$, red dotted curve) overestimates the low-field recovery. Reproducing the experimental field profile requires a substantial contribution from both bands ($\tilde c_h\sim0.5$, solid black curve). While replacing the uniform gap in $R(y)$ with $\Delta_i(\Hfield, T)$ is only a phenomenological approximation deep in the mixed state, this field-dependent constraint independently corroborates our zero-field two-band decomposition ($c_h = \holeweight$).

\subsection{Debye--Callaway model}
To model the lattice thermal conductivity $\kph(T)$ and its pronounced enhancement below $\Tc$ [\FigRef{fig:figure_thermo_and_transport}(d) of the main text], we employ the Debye--Callaway model~\cite{callaway_1959_lattice_thermal_conductivity, Berman1976, Lyu2025}. Assuming Matthiessen's rule, the total phonon relaxation rate $\tau_\mathrm{ph}^{-1}$ is the sum of boundary ($\taubd^{-1}$), point-defect ($\taud^{-1}$), and phonon-electron ($\taueph^{-1}$) scattering rates:
\begin{equation}
    \tau_\mathrm{ph}^{-1} = \taubd^{-1} + \taud^{-1} + \taueph^{-1},
\end{equation}
whose dependencies on the reduced phonon frequency $x = \hbar\omega/(\kb T)$ and reduced temperature $t=T/\Tc$ are summarized in Table~\ref{tab:scattering_rates}.

\begin{table}[htbp]
\centering
\renewcommand{\arraystretch}{1.3}
\begin{tabular}{ll}
\toprule
Mechanism & Scattering rate \\
\midrule
Boundary        & $\taubd^{-1} \propto \mathrm{const}$ \\
Point defect    & $\taud^{-1} \propto x^4 t^4$ \\
Phonon-electron & $\taueph^{-1} \propto x t \, g(x,y)$ \\
\bottomrule
\end{tabular}
\caption{Dominant phonon scattering mechanisms and their functional dependencies on the reduced phonon frequency $x = \hbar\omega/(\kb T)$, reduced gap $y = \bar{\Delta}/(\kb T)$, and reduced temperature $t = T/\Tc$~\cite{tewordtTheoryThermalConductivity1989}.}\label{tab:scattering_rates}
\end{table}

Below $\Tc$, the condensation of electronic states into the superconducting ground state suppresses phonon-electron scattering. According to BRT theory, the ratio of the superconducting- to normal-state phonon-electron scattering rates, $\tauephs^{-1}/\tauephn^{-1} = g(x,y)$, is given by~\cite{bardeenTheoryThermalConductivity1959, tewordtTheoryThermalConductivity1989}
\begin{equation}
    g(x,y) = \frac{1 - e^{-x}}{x}\left[2J_1(x,y) + J_2(x,y)\right],
\end{equation}
where $y=\bar{\Delta}(T)/(\kb T)$ is the reduced gap, and the integrals $J_1(x,y)$ and $J_2(x,y)$ account for phonon scattering by thermally excited Bogoliubov quasiparticles and phonon-induced Cooper pair breaking/recombination (active only for phonon energies $\hbar\omega > 2 \bar{\Delta}$, or $x>2y$), respectively. Above $\Tc$, the vanishing of the gap ($y=0$) yields $g(x,0)=1$, smoothly recovering the normal-state metallic scattering rate. To avoid over-parametrizing the lattice fit with separate electron- and hole-band coupling constants, we evaluate $g(x,y)$ using a single density-of-states-weighted effective gap $\bar{\Delta} = a_e\Delta_e(T) + a_h\Delta_h(T)$, where the gap magnitudes $\Delta_{e,h}(T)$ and Sommerfeld fractions $a_{e,h}$ are fixed directly by our specific heat analysis (Table~\ref{tab:hc_table}). Factoring out the frequency- and temperature-independent boundary scattering rate $\taubd^{-1}$ yields the dimensionless relative scattering rate
\begin{equation}
    r(x,t) \equiv \frac{\tau_\mathrm{ph}^{-1}}{\taubd^{-1}} = 1 + \Gamma_\mathrm{d} x^4 t^4 + \Gamma_\mathrm{ph\text{-}e} x t \, g(x,y),
\end{equation}
where $\Gamma_\mathrm{d}$ and $\Gamma_\mathrm{ph\text{-}e}$ parametrize the relative strengths of point-defect and phonon-electron scattering relative to boundary scattering. Within the Debye--Callaway model, the phonon thermal conductivity is then given by
\begin{equation}
    \kph(T) = \Lambda t^3 \int_0^{\Theta_D(T)/T} \frac{x^4 e^x}{(e^x - 1)^2} \frac{\dd x}{r(x,t)},
\end{equation}
where $\Lambda = (4\pi/3)^{1/3} \kb^{2}\Tc^3\ell_b / [2\pi\hbar a^2\Theta_D^2(T)]$, $a = \SI{7.53}{\angstrom}$ is the conventional cubic lattice constant of \material{}, and $\ell_b$ is the effective phonon boundary mean free path~\cite{tewordtTheoryThermalConductivity1989}.

Because the electronic channels $\ke(T)$ and $\kh(T)$ are completely fixed by specific heat and normal-state magnetotransport, we fit $\kph(T)$ directly to the residual lattice conductivity $\kxx(T) - [\ke(T)+\kh(T)]$. The fit yields a boundary mean free path $\ell_b=\SI{\twobphpathval}{\milli\meter}$, comparable to the physical cross-sectional dimensions of the crystal. We find that phonon-electron scattering strongly dominates the thermal resistance near $\Tc$ ($\Gamma_\mathrm{ph\text{-}e} = \twobgammaephval$), whereas point-defect scattering remains small, $\Gamma_{\mathrm{d}}(T)=1.7[\Theta_D(0)/\Theta_D(T)]^4$ (varying between $0.3 < \Gamma_\mathrm{d} < 1.7$ due to the temperature-dependent Debye temperature $\Theta_D(T)$ extracted from specific heat).


\begin{thebibliography}{44}%
\makeatletter
\providecommand \@ifxundefined [1]{%
 \@ifx{#1\undefined}
}%
\providecommand \@ifnum [1]{%
 \ifnum #1\expandafter \@firstoftwo
 \else \expandafter \@secondoftwo
 \fi
}%
\providecommand \@ifx [1]{%
 \ifx #1\expandafter \@firstoftwo
 \else \expandafter \@secondoftwo
 \fi
}%
\providecommand \natexlab [1]{#1}%
\providecommand \enquote  [1]{``#1''}%
\providecommand \bibnamefont  [1]{#1}%
\providecommand \bibfnamefont [1]{#1}%
\providecommand \citenamefont [1]{#1}%
\providecommand \href@noop [0]{\@secondoftwo}%
\providecommand \href [0]{\begingroup \@sanitize@url \@href}%
\providecommand \@href[1]{\@@startlink{#1}\@@href}%
\providecommand \@@href[1]{\endgroup#1\@@endlink}%
\providecommand \@sanitize@url [0]{\catcode `\\12\catcode `\$12\catcode `\&12\catcode `\#12\catcode `\^12\catcode `\_12\catcode `\%12\relax}%
\providecommand \@@startlink[1]{}%
\providecommand \@@endlink[0]{}%
\providecommand \url  [0]{\begingroup\@sanitize@url \@url }%
\providecommand \@url [1]{\endgroup\@href {#1}{\urlprefix }}%
\providecommand \urlprefix  [0]{URL }%
\providecommand \Eprint [0]{\href }%
\providecommand \doibase [0]{https://doi.org/}%
\providecommand \selectlanguage [0]{\@gobble}%
\providecommand \bibinfo  [0]{\@secondoftwo}%
\providecommand \bibfield  [0]{\@secondoftwo}%
\providecommand \translation [1]{[#1]}%
\providecommand \BibitemOpen [0]{}%
\providecommand \bibitemStop [0]{}%
\providecommand \bibitemNoStop [0]{.\EOS\space}%
\providecommand \EOS [0]{\spacefactor3000\relax}%
\providecommand \BibitemShut  [1]{\csname bibitem#1\endcsname}%
\let\auto@bib@innerbib\@empty
%</preamble>
\bibitem [{\citenamefont {Suhl}\ \emph {et~al.}(1959)\citenamefont {Suhl}, \citenamefont {Matthias},\ and\ \citenamefont {Walker}}]{Suhl1959}%
  \BibitemOpen
  \bibfield  {author} {\bibinfo {author} {\bibfnamefont {H.}~\bibnamefont {Suhl}}, \bibinfo {author} {\bibfnamefont {B.~T.}\ \bibnamefont {Matthias}},\ and\ \bibinfo {author} {\bibfnamefont {L.~R.}\ \bibnamefont {Walker}},\ }\bibfield  {title} {\bibinfo {title} {{Bardeen-Cooper-Schrieffer} theory of superconductivity in the case of overlapping bands},\ }\href {https://doi.org/10.1103/PhysRevLett.3.552} {\bibfield  {journal} {\bibinfo  {journal} {Phys. Rev. Lett.}\ }\textbf {\bibinfo {volume} {3}},\ \bibinfo {pages} {552} (\bibinfo {year} {1959})}\BibitemShut {NoStop}%
\bibitem [{\citenamefont {Moskalenko}(1974)}]{Moskalenko1974}%
  \BibitemOpen
  \bibfield  {author} {\bibinfo {author} {\bibfnamefont {V.~A.}\ \bibnamefont {Moskalenko}},\ }\bibfield  {title} {\bibinfo {title} {The theory of superconductors with overlapping energy bands},\ }\href {https://doi.org/10.1070/PU1974v017n03ABEH004162} {\bibfield  {journal} {\bibinfo  {journal} {Sov. Phys. Usp.}\ }\textbf {\bibinfo {volume} {17}},\ \bibinfo {pages} {450} (\bibinfo {year} {1974})}\BibitemShut {NoStop}%
\bibitem [{\citenamefont {Choi}\ \emph {et~al.}(2002)\citenamefont {Choi}, \citenamefont {Roundy}, \citenamefont {Sun}, \citenamefont {Cohen},\ and\ \citenamefont {Louie}}]{Choi2002_mgb2}%
  \BibitemOpen
  \bibfield  {author} {\bibinfo {author} {\bibfnamefont {H.~J.}\ \bibnamefont {Choi}}, \bibinfo {author} {\bibfnamefont {D.}~\bibnamefont {Roundy}}, \bibinfo {author} {\bibfnamefont {H.}~\bibnamefont {Sun}}, \bibinfo {author} {\bibfnamefont {M.~L.}\ \bibnamefont {Cohen}},\ and\ \bibinfo {author} {\bibfnamefont {S.~G.}\ \bibnamefont {Louie}},\ }\bibfield  {title} {\bibinfo {title} {The origin of the anomalous superconducting properties of {MgB}$_2$},\ }\href {https://doi.org/10.1038/nature00898} {\bibfield  {journal} {\bibinfo  {journal} {Nature}\ }\textbf {\bibinfo {volume} {418}},\ \bibinfo {pages} {758} (\bibinfo {year} {2002})}\BibitemShut {NoStop}%
\bibitem [{\citenamefont {Souma}\ \emph {et~al.}(2003)\citenamefont {Souma}, \citenamefont {Machida}, \citenamefont {Sato}, \citenamefont {Takahashi}, \citenamefont {Matsui}, \citenamefont {Wang}, \citenamefont {Ding}, \citenamefont {Kaminski}, \citenamefont {Campuzano}, \citenamefont {Sasaki},\ and\ \citenamefont {Kadowaki}}]{Souma2003}%
  \BibitemOpen
  \bibfield  {author} {\bibinfo {author} {\bibfnamefont {S.}~\bibnamefont {Souma}}, \bibinfo {author} {\bibfnamefont {Y.}~\bibnamefont {Machida}}, \bibinfo {author} {\bibfnamefont {T.}~\bibnamefont {Sato}}, \bibinfo {author} {\bibfnamefont {T.}~\bibnamefont {Takahashi}}, \bibinfo {author} {\bibfnamefont {H.}~\bibnamefont {Matsui}}, \bibinfo {author} {\bibfnamefont {S.-C.}\ \bibnamefont {Wang}}, \bibinfo {author} {\bibfnamefont {H.}~\bibnamefont {Ding}}, \bibinfo {author} {\bibfnamefont {A.}~\bibnamefont {Kaminski}}, \bibinfo {author} {\bibfnamefont {J.~C.}\ \bibnamefont {Campuzano}}, \bibinfo {author} {\bibfnamefont {S.}~\bibnamefont {Sasaki}},\ and\ \bibinfo {author} {\bibfnamefont {K.}~\bibnamefont {Kadowaki}},\ }\bibfield  {title} {\bibinfo {title} {The origin of multiple superconducting gaps in {MgB}$_2$},\ }\href {https://doi.org/10.1038/nature01619} {\bibfield  {journal} {\bibinfo  {journal} {Nature}\ }\textbf {\bibinfo {volume} {423}},\ \bibinfo {pages} {65} (\bibinfo {year} {2003})}\BibitemShut
  {NoStop}%
\bibitem [{\citenamefont {Yokoya}\ \emph {et~al.}(2001)\citenamefont {Yokoya}, \citenamefont {Kiss}, \citenamefont {Chainani}, \citenamefont {Shin}, \citenamefont {Nohara},\ and\ \citenamefont {Takagi}}]{Yokoya2001}%
  \BibitemOpen
  \bibfield  {author} {\bibinfo {author} {\bibfnamefont {T.}~\bibnamefont {Yokoya}}, \bibinfo {author} {\bibfnamefont {T.}~\bibnamefont {Kiss}}, \bibinfo {author} {\bibfnamefont {A.}~\bibnamefont {Chainani}}, \bibinfo {author} {\bibfnamefont {S.}~\bibnamefont {Shin}}, \bibinfo {author} {\bibfnamefont {M.}~\bibnamefont {Nohara}},\ and\ \bibinfo {author} {\bibfnamefont {H.}~\bibnamefont {Takagi}},\ }\bibfield  {title} {\bibinfo {title} {{Fermi} surface sheet-dependent superconductivity in {2$H$-NbSe}$_2$},\ }\href {https://doi.org/10.1126/science.1065068} {\bibfield  {journal} {\bibinfo  {journal} {Science}\ }\textbf {\bibinfo {volume} {294}},\ \bibinfo {pages} {2518} (\bibinfo {year} {2001})}\BibitemShut {NoStop}%
\bibitem [{\citenamefont {Ding}\ \emph {et~al.}(2008)\citenamefont {Ding}, \citenamefont {Nakayama}, \citenamefont {Richard}, \citenamefont {Souma}, \citenamefont {Sato}, \citenamefont {Takahashi}, \citenamefont {Neupane}, \citenamefont {Xu}, \citenamefont {Pan}, \citenamefont {Federov}, \citenamefont {Wang}, \citenamefont {Dai}, \citenamefont {Fang}, \citenamefont {Chen}, \citenamefont {Luo},\ and\ \citenamefont {Wang}}]{Ding2008}%
  \BibitemOpen
  \bibfield  {author} {\bibinfo {author} {\bibfnamefont {H.}~\bibnamefont {Ding}}, \bibinfo {author} {\bibfnamefont {K.}~\bibnamefont {Nakayama}}, \bibinfo {author} {\bibfnamefont {P.}~\bibnamefont {Richard}}, \bibinfo {author} {\bibfnamefont {S.}~\bibnamefont {Souma}}, \bibinfo {author} {\bibfnamefont {T.}~\bibnamefont {Sato}}, \bibinfo {author} {\bibfnamefont {T.}~\bibnamefont {Takahashi}}, \bibinfo {author} {\bibfnamefont {M.}~\bibnamefont {Neupane}}, \bibinfo {author} {\bibfnamefont {Y.-M.}\ \bibnamefont {Xu}}, \bibinfo {author} {\bibfnamefont {Z.-H.}\ \bibnamefont {Pan}}, \bibinfo {author} {\bibfnamefont {A.~V.}\ \bibnamefont {Federov}}, \bibinfo {author} {\bibfnamefont {Z.}~\bibnamefont {Wang}}, \bibinfo {author} {\bibfnamefont {X.}~\bibnamefont {Dai}}, \bibinfo {author} {\bibfnamefont {Z.}~\bibnamefont {Fang}}, \bibinfo {author} {\bibfnamefont {G.~F.}\ \bibnamefont {Chen}}, \bibinfo {author} {\bibfnamefont {J.~L.}\ \bibnamefont {Luo}},\ and\ \bibinfo {author} {\bibfnamefont {N.~L.}\ \bibnamefont
  {Wang}},\ }\bibfield  {title} {\bibinfo {title} {Electronic dilation and superconducting gaps in {Ba}$_{0.6}${K}$_{0.4}${Fe}$_2${As}$_2$},\ }\href {https://doi.org/10.1209/0295-5075/83/47001} {\bibfield  {journal} {\bibinfo  {journal} {EPL}\ }\textbf {\bibinfo {volume} {83}},\ \bibinfo {pages} {47001} (\bibinfo {year} {2008})}\BibitemShut {NoStop}%
\bibitem [{\citenamefont {Steglich}\ \emph {et~al.}(1979)\citenamefont {Steglich}, \citenamefont {Aarts}, \citenamefont {Bredl}, \citenamefont {Lieke}, \citenamefont {Meschede}, \citenamefont {Franz},\ and\ \citenamefont {Sch{\"a}fer}}]{Steglich1979}%
  \BibitemOpen
  \bibfield  {author} {\bibinfo {author} {\bibfnamefont {F.}~\bibnamefont {Steglich}}, \bibinfo {author} {\bibfnamefont {J.}~\bibnamefont {Aarts}}, \bibinfo {author} {\bibfnamefont {C.~D.}\ \bibnamefont {Bredl}}, \bibinfo {author} {\bibfnamefont {W.}~\bibnamefont {Lieke}}, \bibinfo {author} {\bibfnamefont {D.}~\bibnamefont {Meschede}}, \bibinfo {author} {\bibfnamefont {W.}~\bibnamefont {Franz}},\ and\ \bibinfo {author} {\bibfnamefont {H.}~\bibnamefont {Sch{\"a}fer}},\ }\bibfield  {title} {\bibinfo {title} {Superconductivity in the presence of strong pauli paramagnetism: {$\mathrm{CeCu_{2}Si_{2}}$}},\ }\href {https://doi.org/10.1103/PhysRevLett.43.1892} {\bibfield  {journal} {\bibinfo  {journal} {Phys. Rev. Lett.}\ }\textbf {\bibinfo {volume} {43}},\ \bibinfo {pages} {1892} (\bibinfo {year} {1979})}\BibitemShut {NoStop}%
\bibitem [{\citenamefont {Pfleiderer}(2009)}]{Pfleiderer2009}%
  \BibitemOpen
  \bibfield  {author} {\bibinfo {author} {\bibfnamefont {C.}~\bibnamefont {Pfleiderer}},\ }\bibfield  {title} {\bibinfo {title} {Superconducting phases of $f$-electron compounds},\ }\href {https://doi.org/10.1103/RevModPhys.81.1551} {\bibfield  {journal} {\bibinfo  {journal} {Rev. Mod. Phys.}\ }\textbf {\bibinfo {volume} {81}},\ \bibinfo {pages} {1551} (\bibinfo {year} {2009})}\BibitemShut {NoStop}%
\bibitem [{\citenamefont {Seyfarth}\ \emph {et~al.}(2008)\citenamefont {Seyfarth}, \citenamefont {Brison}, \citenamefont {Knebel}, \citenamefont {Aoki}, \citenamefont {Lapertot},\ and\ \citenamefont {Flouquet}}]{Seyfarth2008}%
  \BibitemOpen
  \bibfield  {author} {\bibinfo {author} {\bibfnamefont {G.}~\bibnamefont {Seyfarth}}, \bibinfo {author} {\bibfnamefont {J.~P.}\ \bibnamefont {Brison}}, \bibinfo {author} {\bibfnamefont {G.}~\bibnamefont {Knebel}}, \bibinfo {author} {\bibfnamefont {D.}~\bibnamefont {Aoki}}, \bibinfo {author} {\bibfnamefont {G.}~\bibnamefont {Lapertot}},\ and\ \bibinfo {author} {\bibfnamefont {J.}~\bibnamefont {Flouquet}},\ }\bibfield  {title} {\bibinfo {title} {Multigap superconductivity in the heavy-fermion system {$\mathrm{CeCoIn_{5}}$}},\ }\href {https://doi.org/10.1103/PhysRevLett.101.046401} {\bibfield  {journal} {\bibinfo  {journal} {Phys. Rev. Lett.}\ }\textbf {\bibinfo {volume} {101}},\ \bibinfo {pages} {046401} (\bibinfo {year} {2008})}\BibitemShut {NoStop}%
\bibitem [{\citenamefont {Allan}\ \emph {et~al.}(2013)\citenamefont {Allan}, \citenamefont {Massee}, \citenamefont {Morr}, \citenamefont {Van~Dyke}, \citenamefont {Rost}, \citenamefont {Mackenzie}, \citenamefont {Petrovic},\ and\ \citenamefont {Davis}}]{Allan2013}%
  \BibitemOpen
  \bibfield  {author} {\bibinfo {author} {\bibfnamefont {M.~P.}\ \bibnamefont {Allan}}, \bibinfo {author} {\bibfnamefont {F.}~\bibnamefont {Massee}}, \bibinfo {author} {\bibfnamefont {D.~K.}\ \bibnamefont {Morr}}, \bibinfo {author} {\bibfnamefont {J.}~\bibnamefont {Van~Dyke}}, \bibinfo {author} {\bibfnamefont {A.~W.}\ \bibnamefont {Rost}}, \bibinfo {author} {\bibfnamefont {A.~P.}\ \bibnamefont {Mackenzie}}, \bibinfo {author} {\bibfnamefont {C.}~\bibnamefont {Petrovic}},\ and\ \bibinfo {author} {\bibfnamefont {J.~C.}\ \bibnamefont {Davis}},\ }\bibfield  {title} {\bibinfo {title} {Imaging cooper pairing of heavy fermions in {$\mathrm{CeCoIn_{5}}$}},\ }\href {https://doi.org/10.1038/nphys2671} {\bibfield  {journal} {\bibinfo  {journal} {Nat. Phys.}\ }\textbf {\bibinfo {volume} {9}},\ \bibinfo {pages} {468} (\bibinfo {year} {2013})}\BibitemShut {NoStop}%
\bibitem [{\citenamefont {Kittaka}\ \emph {et~al.}(2014)\citenamefont {Kittaka}, \citenamefont {Aoki}, \citenamefont {Shimura}, \citenamefont {Sakakibara}, \citenamefont {Seiro}, \citenamefont {Geibel}, \citenamefont {Steglich}, \citenamefont {Ikeda},\ and\ \citenamefont {Machida}}]{Kittaka2014}%
  \BibitemOpen
  \bibfield  {author} {\bibinfo {author} {\bibfnamefont {S.}~\bibnamefont {Kittaka}}, \bibinfo {author} {\bibfnamefont {Y.}~\bibnamefont {Aoki}}, \bibinfo {author} {\bibfnamefont {Y.}~\bibnamefont {Shimura}}, \bibinfo {author} {\bibfnamefont {T.}~\bibnamefont {Sakakibara}}, \bibinfo {author} {\bibfnamefont {S.}~\bibnamefont {Seiro}}, \bibinfo {author} {\bibfnamefont {C.}~\bibnamefont {Geibel}}, \bibinfo {author} {\bibfnamefont {F.}~\bibnamefont {Steglich}}, \bibinfo {author} {\bibfnamefont {H.}~\bibnamefont {Ikeda}},\ and\ \bibinfo {author} {\bibfnamefont {K.}~\bibnamefont {Machida}},\ }\bibfield  {title} {\bibinfo {title} {Multiband superconductivity with unexpected deficiency of nodal quasiparticles in {$\mathrm{CeCu_{2}Si_{2}}$}},\ }\href {https://doi.org/10.1103/PhysRevLett.112.067002} {\bibfield  {journal} {\bibinfo  {journal} {Phys. Rev. Lett.}\ }\textbf {\bibinfo {volume} {112}},\ \bibinfo {pages} {067002} (\bibinfo {year} {2014})}\BibitemShut {NoStop}%
\bibitem [{\citenamefont {Gallagher}\ \emph {et~al.}(2016)\citenamefont {Gallagher}, \citenamefont {Chen}, \citenamefont {Moir}, \citenamefont {Cary}, \citenamefont {Kametani}, \citenamefont {Graf}, \citenamefont {Albrecht-Schmitt}, \citenamefont {Riggs}, \citenamefont {Shekhter},\ and\ \citenamefont {Baumbach}}]{Gallagher2016}%
  \BibitemOpen
  \bibfield  {author} {\bibinfo {author} {\bibfnamefont {A.}~\bibnamefont {Gallagher}}, \bibinfo {author} {\bibfnamefont {K.-W.}\ \bibnamefont {Chen}}, \bibinfo {author} {\bibfnamefont {C.~M.}\ \bibnamefont {Moir}}, \bibinfo {author} {\bibfnamefont {S.~K.}\ \bibnamefont {Cary}}, \bibinfo {author} {\bibfnamefont {F.}~\bibnamefont {Kametani}}, \bibinfo {author} {\bibfnamefont {D.}~\bibnamefont {Graf}}, \bibinfo {author} {\bibfnamefont {T.~E.}\ \bibnamefont {Albrecht-Schmitt}}, \bibinfo {author} {\bibfnamefont {S.~C.}\ \bibnamefont {Riggs}}, \bibinfo {author} {\bibfnamefont {A.}~\bibnamefont {Shekhter}},\ and\ \bibinfo {author} {\bibfnamefont {R.~E.}\ \bibnamefont {Baumbach}},\ }\bibfield  {title} {\bibinfo {title} {Unfolding the physics of {$\mathrm{URu_{2}Si_{2}}$} through silicon to phosphorus substitution},\ }\href {https://doi.org/10.1038/ncomms10712} {\bibfield  {journal} {\bibinfo  {journal} {Nat. Commun.}\ }\textbf {\bibinfo {volume} {7}},\ \bibinfo {pages} {10712} (\bibinfo {year} {2016})}\BibitemShut
  {NoStop}%
\bibitem [{\citenamefont {Hirschfeld}\ \emph {et~al.}(2011)\citenamefont {Hirschfeld}, \citenamefont {Korshunov},\ and\ \citenamefont {Mazin}}]{Hirschfeld2011}%
  \BibitemOpen
  \bibfield  {author} {\bibinfo {author} {\bibfnamefont {P.~J.}\ \bibnamefont {Hirschfeld}}, \bibinfo {author} {\bibfnamefont {M.~M.}\ \bibnamefont {Korshunov}},\ and\ \bibinfo {author} {\bibfnamefont {I.~I.}\ \bibnamefont {Mazin}},\ }\bibfield  {title} {\bibinfo {title} {Gap symmetry and structure of {Fe}-based superconductors},\ }\href {https://doi.org/10.1088/0034-4885/74/12/124508} {\bibfield  {journal} {\bibinfo  {journal} {Rep. Prog. Phys.}\ }\textbf {\bibinfo {volume} {74}},\ \bibinfo {pages} {124508} (\bibinfo {year} {2011})}\BibitemShut {NoStop}%
\bibitem [{\citenamefont {Matsuda}\ \emph {et~al.}(2006)\citenamefont {Matsuda}, \citenamefont {Izawa},\ and\ \citenamefont {Vekhter}}]{Matsuda2006}%
  \BibitemOpen
  \bibfield  {author} {\bibinfo {author} {\bibfnamefont {Y.}~\bibnamefont {Matsuda}}, \bibinfo {author} {\bibfnamefont {K.}~\bibnamefont {Izawa}},\ and\ \bibinfo {author} {\bibfnamefont {I.}~\bibnamefont {Vekhter}},\ }\bibfield  {title} {\bibinfo {title} {Nodal structure of unconventional superconductors probed by the angle resolved thermal transport measurements},\ }\href {https://doi.org/10.1088/0953-8984/18/44/R01} {\bibfield  {journal} {\bibinfo  {journal} {J. Phys.: Condens. Matter}\ }\textbf {\bibinfo {volume} {18}},\ \bibinfo {pages} {R705} (\bibinfo {year} {2006})}\BibitemShut {NoStop}%
\bibitem [{\citenamefont {Boaknin}\ \emph {et~al.}(2003)\citenamefont {Boaknin}, \citenamefont {Tanatar} \emph {et~al.}}]{boakninHeatConductionVortex2003}%
  \BibitemOpen
  \bibfield  {author} {\bibinfo {author} {\bibfnamefont {E.}~\bibnamefont {Boaknin}}, \bibinfo {author} {\bibfnamefont {M.~A.}\ \bibnamefont {Tanatar}}, \emph {et~al.},\ }\bibfield  {title} {\bibinfo {title} {Heat conduction in the vortex state of {NbSe}$_2$: {Evidence} for multiband superconductivity},\ }\href {https://doi.org/10.1103/PhysRevLett.90.117003} {\bibfield  {journal} {\bibinfo  {journal} {Phys. Rev. Lett.}\ }\textbf {\bibinfo {volume} {90}},\ \bibinfo {pages} {117003} (\bibinfo {year} {2003})}\BibitemShut {NoStop}%
\bibitem [{\citenamefont {Bouquet}\ \emph {et~al.}(2001)\citenamefont {Bouquet}, \citenamefont {Fisher}, \citenamefont {Phillips}, \citenamefont {Hinks},\ and\ \citenamefont {Jorgensen}}]{Bouquet2001}%
  \BibitemOpen
  \bibfield  {author} {\bibinfo {author} {\bibfnamefont {F.}~\bibnamefont {Bouquet}}, \bibinfo {author} {\bibfnamefont {R.~A.}\ \bibnamefont {Fisher}}, \bibinfo {author} {\bibfnamefont {N.~E.}\ \bibnamefont {Phillips}}, \bibinfo {author} {\bibfnamefont {D.~G.}\ \bibnamefont {Hinks}},\ and\ \bibinfo {author} {\bibfnamefont {J.~D.}\ \bibnamefont {Jorgensen}},\ }\bibfield  {title} {\bibinfo {title} {Specific heat of {Mg$^{11}$B$_2$}: {Evidence} for a second energy gap},\ }\href {https://doi.org/10.1103/PhysRevLett.87.047001} {\bibfield  {journal} {\bibinfo  {journal} {Phys. Rev. Lett.}\ }\textbf {\bibinfo {volume} {87}},\ \bibinfo {pages} {047001} (\bibinfo {year} {2001})}\BibitemShut {NoStop}%
\bibitem [{\citenamefont {Prozorov}\ and\ \citenamefont {Giannetta}(2006)}]{Prozorov2006}%
  \BibitemOpen
  \bibfield  {author} {\bibinfo {author} {\bibfnamefont {R.}~\bibnamefont {Prozorov}}\ and\ \bibinfo {author} {\bibfnamefont {R.~W.}\ \bibnamefont {Giannetta}},\ }\bibfield  {title} {\bibinfo {title} {Magnetic penetration depth in unconventional superconductors},\ }\href {https://doi.org/10.1088/0953-2048/19/8/R01} {\bibfield  {journal} {\bibinfo  {journal} {Supercond. Sci. Technol.}\ }\textbf {\bibinfo {volume} {19}},\ \bibinfo {pages} {R41} (\bibinfo {year} {2006})}\BibitemShut {NoStop}%
\bibitem [{\citenamefont {Shakeripour}\ \emph {et~al.}(2009)\citenamefont {Shakeripour}, \citenamefont {Petrovic},\ and\ \citenamefont {Taillefer}}]{Shakeripour2009}%
  \BibitemOpen
  \bibfield  {author} {\bibinfo {author} {\bibfnamefont {H.}~\bibnamefont {Shakeripour}}, \bibinfo {author} {\bibfnamefont {C.}~\bibnamefont {Petrovic}},\ and\ \bibinfo {author} {\bibfnamefont {L.}~\bibnamefont {Taillefer}},\ }\bibfield  {title} {\bibinfo {title} {Heat transport as a probe of superconducting gap structure},\ }\href {https://doi.org/10.1088/1367-2630/11/5/055065} {\bibfield  {journal} {\bibinfo  {journal} {New J. Phys.}\ }\textbf {\bibinfo {volume} {11}},\ \bibinfo {pages} {055065} (\bibinfo {year} {2009})}\BibitemShut {NoStop}%
\bibitem [{\citenamefont {Johnston}(2013)}]{Johnston2013}%
  \BibitemOpen
  \bibfield  {author} {\bibinfo {author} {\bibfnamefont {D.~C.}\ \bibnamefont {Johnston}},\ }\bibfield  {title} {\bibinfo {title} {Elaboration of the {$\alpha$}-model derived from the {BCS} theory of superconductivity},\ }\href {https://doi.org/10.1088/0953-2048/26/11/115011} {\bibfield  {journal} {\bibinfo  {journal} {Supercond. Sci. Technol.}\ }\textbf {\bibinfo {volume} {26}},\ \bibinfo {pages} {115011} (\bibinfo {year} {2013})}\BibitemShut {NoStop}%
\bibitem [{\citenamefont {Fletcher}\ \emph {et~al.}(2007)\citenamefont {Fletcher}, \citenamefont {Carrington}, \citenamefont {Diener}, \citenamefont {Rodi{\`e}re}, \citenamefont {Brison}, \citenamefont {Prozorov}, \citenamefont {Olheiser},\ and\ \citenamefont {Giannetta}}]{Fletcher2007}%
  \BibitemOpen
  \bibfield  {author} {\bibinfo {author} {\bibfnamefont {J.~D.}\ \bibnamefont {Fletcher}}, \bibinfo {author} {\bibfnamefont {A.}~\bibnamefont {Carrington}}, \bibinfo {author} {\bibfnamefont {P.}~\bibnamefont {Diener}}, \bibinfo {author} {\bibfnamefont {P.}~\bibnamefont {Rodi{\`e}re}}, \bibinfo {author} {\bibfnamefont {J.~P.}\ \bibnamefont {Brison}}, \bibinfo {author} {\bibfnamefont {R.}~\bibnamefont {Prozorov}}, \bibinfo {author} {\bibfnamefont {T.}~\bibnamefont {Olheiser}},\ and\ \bibinfo {author} {\bibfnamefont {R.~W.}\ \bibnamefont {Giannetta}},\ }\bibfield  {title} {\bibinfo {title} {Penetration depth study of superconducting gap structure of {2$H$-NbSe}$_2$},\ }\href {https://doi.org/10.1103/PhysRevLett.98.057003} {\bibfield  {journal} {\bibinfo  {journal} {Phys. Rev. Lett.}\ }\textbf {\bibinfo {volume} {98}},\ \bibinfo {pages} {057003} (\bibinfo {year} {2007})}\BibitemShut {NoStop}%
\bibitem [{\citenamefont {Zehetmayer}(2013)}]{Zehetmayer2013}%
  \BibitemOpen
  \bibfield  {author} {\bibinfo {author} {\bibfnamefont {M.}~\bibnamefont {Zehetmayer}},\ }\bibfield  {title} {\bibinfo {title} {A review of two-band superconductivity: materials and effects on the thermodynamic and reversible mixed-state properties},\ }\href {https://doi.org/10.1088/0953-2048/26/4/043001} {\bibfield  {journal} {\bibinfo  {journal} {Supercond. Sci. Technol.}\ }\textbf {\bibinfo {volume} {26}},\ \bibinfo {pages} {043001} (\bibinfo {year} {2013})}\BibitemShut {NoStop}%
\bibitem [{\citenamefont {Onose}\ \emph {et~al.}(2007)\citenamefont {Onose}, \citenamefont {Ong},\ and\ \citenamefont {Petrovic}}]{onoseLorenzNumberCeCoIn52007}%
  \BibitemOpen
  \bibfield  {author} {\bibinfo {author} {\bibfnamefont {Y.}~\bibnamefont {Onose}}, \bibinfo {author} {\bibfnamefont {N.~P.}\ \bibnamefont {Ong}},\ and\ \bibinfo {author} {\bibfnamefont {C.}~\bibnamefont {Petrovic}},\ }\bibfield  {title} {\bibinfo {title} {The {Lorenz} number in {CeCoIn}$_5$ inferred from the thermal and charge {Hall} currents},\ }\href {https://doi.org/10.1209/0295-5075/80/37005} {\bibfield  {journal} {\bibinfo  {journal} {EPL}\ }\textbf {\bibinfo {volume} {80}},\ \bibinfo {pages} {37005} (\bibinfo {year} {2007})}\BibitemShut {NoStop}%
\bibitem [{\citenamefont {Checkelsky}\ \emph {et~al.}(2012)\citenamefont {Checkelsky} \emph {et~al.}}]{checkelsky2012thermal}%
  \BibitemOpen
  \bibfield  {author} {\bibinfo {author} {\bibfnamefont {J.~G.}\ \bibnamefont {Checkelsky}} \emph {et~al.},\ }\bibfield  {title} {\bibinfo {title} {Thermal {Hall} conductivity as a probe of gap structure in multiband superconductors: {The} case of {{Ba}}$_{1-x}${{K}}$_x${{Fe}}$_2${{As}}$_2$},\ }\href {https://doi.org/10.1103/PhysRevB.86.180502} {\bibfield  {journal} {\bibinfo  {journal} {Phys. Rev. B}\ }\textbf {\bibinfo {volume} {86}},\ \bibinfo {pages} {180502(R)} (\bibinfo {year} {2012})}\BibitemShut {NoStop}%
\bibitem [{\citenamefont {Bardeen}\ \emph {et~al.}(1959)\citenamefont {Bardeen}, \citenamefont {Rickayzen},\ and\ \citenamefont {Tewordt}}]{bardeenTheoryThermalConductivity1959}%
  \BibitemOpen
  \bibfield  {author} {\bibinfo {author} {\bibfnamefont {J.}~\bibnamefont {Bardeen}}, \bibinfo {author} {\bibfnamefont {G.}~\bibnamefont {Rickayzen}},\ and\ \bibinfo {author} {\bibfnamefont {L.}~\bibnamefont {Tewordt}},\ }\bibfield  {title} {\bibinfo {title} {Theory of the thermal conductivity of superconductors},\ }\href {https://doi.org/10.1103/PhysRev.113.982} {\bibfield  {journal} {\bibinfo  {journal} {Phys. Rev.}\ }\textbf {\bibinfo {volume} {113}},\ \bibinfo {pages} {982} (\bibinfo {year} {1959})}\BibitemShut {NoStop}%
\bibitem [{\citenamefont {Tinkham}(2004)}]{tinkham2004introduction}%
  \BibitemOpen
  \bibfield  {author} {\bibinfo {author} {\bibfnamefont {M.}~\bibnamefont {Tinkham}},\ }\href@noop {} {\emph {\bibinfo {title} {Introduction to Superconductivity}}},\ \bibinfo {edition} {2nd}\ ed.\ (\bibinfo  {publisher} {Dover Publications},\ \bibinfo {address} {Mineola, NY},\ \bibinfo {year} {2004})\BibitemShut {NoStop}%
\bibitem [{\citenamefont {Hedo}\ \emph {et~al.}(1995)\citenamefont {Hedo}, \citenamefont {Inada}, \citenamefont {Ishida}, \citenamefont {Yamamoto}, \citenamefont {Haga}, \citenamefont {{\=O}nuki}, \citenamefont {Higuchi},\ and\ \citenamefont {Hasegawa}}]{Hedo1995CeRu2}%
  \BibitemOpen
  \bibfield  {author} {\bibinfo {author} {\bibfnamefont {M.}~\bibnamefont {Hedo}}, \bibinfo {author} {\bibfnamefont {Y.}~\bibnamefont {Inada}}, \bibinfo {author} {\bibfnamefont {T.}~\bibnamefont {Ishida}}, \bibinfo {author} {\bibfnamefont {E.}~\bibnamefont {Yamamoto}}, \bibinfo {author} {\bibfnamefont {Y.}~\bibnamefont {Haga}}, \bibinfo {author} {\bibfnamefont {Y.}~\bibnamefont {{\=O}nuki}}, \bibinfo {author} {\bibfnamefont {M.}~\bibnamefont {Higuchi}},\ and\ \bibinfo {author} {\bibfnamefont {A.}~\bibnamefont {Hasegawa}},\ }\bibfield  {title} {\bibinfo {title} {de {Haas}--van {Alphen} oscillation in both normal and superconducting mixed states of {CeRu}$_2$},\ }\href {https://doi.org/10.1143/JPSJ.64.4535} {\bibfield  {journal} {\bibinfo  {journal} {J. Phys. Soc. Jpn.}\ }\textbf {\bibinfo {volume} {64}},\ \bibinfo {pages} {4535} (\bibinfo {year} {1995})}\BibitemShut {NoStop}%
\bibitem [{\citenamefont {Mielke}\ \emph {et~al.}(2022)\citenamefont {Mielke}, \citenamefont {Liu} \emph {et~al.}}]{mielkeiiiLocalSpectroscopicEvidence2022a}%
  \BibitemOpen
  \bibfield  {author} {\bibinfo {author} {\bibfnamefont {C.}~\bibnamefont {Mielke}, \bibfnamefont {III}}, \bibinfo {author} {\bibfnamefont {H.}~\bibnamefont {Liu}}, \emph {et~al.},\ }\bibfield  {title} {\bibinfo {title} {Local spectroscopic evidence for a nodeless magnetic kagome superconductor {CeRu}$_2$},\ }\href {https://doi.org/10.1088/1361-648X/ac9813} {\bibfield  {journal} {\bibinfo  {journal} {J. Phys.: Condens. Matter}\ }\textbf {\bibinfo {volume} {34}},\ \bibinfo {pages} {485601} (\bibinfo {year} {2022})}\BibitemShut {NoStop}%
\bibitem [{\citenamefont {Kiss}\ \emph {et~al.}(2005)\citenamefont {Kiss}, \citenamefont {Kanetaka}, \citenamefont {Yokoya}, \citenamefont {Shimojima}, \citenamefont {Kanai}, \citenamefont {Shin}, \citenamefont {{\=O}nuki}, \citenamefont {Togashi}, \citenamefont {Zhang}, \citenamefont {Chen},\ and\ \citenamefont {Watanabe}}]{kissPhotoemissionSpectroscopicEvidence2005}%
  \BibitemOpen
  \bibfield  {author} {\bibinfo {author} {\bibfnamefont {T.}~\bibnamefont {Kiss}}, \bibinfo {author} {\bibfnamefont {F.}~\bibnamefont {Kanetaka}}, \bibinfo {author} {\bibfnamefont {T.}~\bibnamefont {Yokoya}}, \bibinfo {author} {\bibfnamefont {T.}~\bibnamefont {Shimojima}}, \bibinfo {author} {\bibfnamefont {K.}~\bibnamefont {Kanai}}, \bibinfo {author} {\bibfnamefont {S.}~\bibnamefont {Shin}}, \bibinfo {author} {\bibfnamefont {Y.}~\bibnamefont {{\=O}nuki}}, \bibinfo {author} {\bibfnamefont {T.}~\bibnamefont {Togashi}}, \bibinfo {author} {\bibfnamefont {C.}~\bibnamefont {Zhang}}, \bibinfo {author} {\bibfnamefont {C.~T.}\ \bibnamefont {Chen}},\ and\ \bibinfo {author} {\bibfnamefont {S.}~\bibnamefont {Watanabe}},\ }\bibfield  {title} {\bibinfo {title} {Photoemission spectroscopic evidence of gap anisotropy in an $f$-electron superconductor},\ }\href {https://doi.org/10.1103/PhysRevLett.94.057001} {\bibfield  {journal} {\bibinfo  {journal} {Phys. Rev. Lett.}\ }\textbf {\bibinfo {volume} {94}},\ \bibinfo {pages}
  {057001} (\bibinfo {year} {2005})}\BibitemShut {NoStop}%
\bibitem [{\citenamefont {Kittaka}\ \emph {et~al.}(2013)\citenamefont {Kittaka}, \citenamefont {Sakakibara} \emph {et~al.}}]{kittakaVerificationAnisotropicWave2013}%
  \BibitemOpen
  \bibfield  {author} {\bibinfo {author} {\bibfnamefont {S.}~\bibnamefont {Kittaka}}, \bibinfo {author} {\bibfnamefont {T.}~\bibnamefont {Sakakibara}}, \emph {et~al.},\ }\bibfield  {title} {\bibinfo {title} {Verification of anisotropic $s$-wave superconducting gap structure in {CeRu}$_2$ from low-temperature field-angle-resolved specific heat measurements},\ }\href {https://doi.org/10.7566/JPSJ.82.123706} {\bibfield  {journal} {\bibinfo  {journal} {J. Phys. Soc. Jpn.}\ }\textbf {\bibinfo {volume} {82}},\ \bibinfo {pages} {123706} (\bibinfo {year} {2013})}\BibitemShut {NoStop}%
\bibitem [{\citenamefont {Gerguri}\ \emph {et~al.}(2026)\citenamefont {Gerguri}, \citenamefont {Das} \emph {et~al.}}]{gerguriDistinctUniaxialStress2026}%
  \BibitemOpen
  \bibfield  {author} {\bibinfo {author} {\bibfnamefont {O.}~\bibnamefont {Gerguri}}, \bibinfo {author} {\bibfnamefont {D.}~\bibnamefont {Das}}, \emph {et~al.},\ }\bibfield  {title} {\bibinfo {title} {Distinct uniaxial stress and pressure fingerprint of superconductivity in the {3D} kagome lattice compound {CeRu}$_2$},\ }\href {https://doi.org/10.1038/s42005-026-02553-3} {\bibfield  {journal} {\bibinfo  {journal} {Commun. Phys.}\ }\textbf {\bibinfo {volume} {9}},\ \bibinfo {pages} {122} (\bibinfo {year} {2026})}\BibitemShut {NoStop}%
\bibitem [{\citenamefont {Higuchi}\ and\ \citenamefont {Hasegawa}(1996)}]{Higuchi1996}%
  \BibitemOpen
  \bibfield  {author} {\bibinfo {author} {\bibfnamefont {M.}~\bibnamefont {Higuchi}}\ and\ \bibinfo {author} {\bibfnamefont {A.}~\bibnamefont {Hasegawa}},\ }\bibfield  {title} {\bibinfo {title} {{Fermi} surface of {CeRu}$_2$ within local-density functional theory},\ }\href {https://doi.org/10.1143/JPSJ.65.1302} {\bibfield  {journal} {\bibinfo  {journal} {J. Phys. Soc. Jpn.}\ }\textbf {\bibinfo {volume} {65}},\ \bibinfo {pages} {1302} (\bibinfo {year} {1996})}\BibitemShut {NoStop}%
\bibitem [{\citenamefont {Huxley}\ \emph {et~al.}(1993)\citenamefont {Huxley}, \citenamefont {Paulsen}, \citenamefont {Laborde}, \citenamefont {Tholence}, \citenamefont {Sanchez}, \citenamefont {Junod},\ and\ \citenamefont {Calemczuk}}]{Huxley_1993_CeRu2}%
  \BibitemOpen
  \bibfield  {author} {\bibinfo {author} {\bibfnamefont {A.~D.}\ \bibnamefont {Huxley}}, \bibinfo {author} {\bibfnamefont {C.}~\bibnamefont {Paulsen}}, \bibinfo {author} {\bibfnamefont {O.}~\bibnamefont {Laborde}}, \bibinfo {author} {\bibfnamefont {J.~L.}\ \bibnamefont {Tholence}}, \bibinfo {author} {\bibfnamefont {D.}~\bibnamefont {Sanchez}}, \bibinfo {author} {\bibfnamefont {A.}~\bibnamefont {Junod}},\ and\ \bibinfo {author} {\bibfnamefont {R.}~\bibnamefont {Calemczuk}},\ }\bibfield  {title} {\bibinfo {title} {Flux pinning, specific heat and magnetic properties of the {Laves} phase superconductor {CeRu}$_2$},\ }\href {https://doi.org/10.1088/0953-8984/5/41/018} {\bibfield  {journal} {\bibinfo  {journal} {J. Phys.: Condens. Matter}\ }\textbf {\bibinfo {volume} {5}},\ \bibinfo {pages} {7709} (\bibinfo {year} {1993})}\BibitemShut {NoStop}%
\bibitem [{\citenamefont {Hedo}\ \emph {et~al.}(1998)\citenamefont {Hedo}, \citenamefont {Inada}, \citenamefont {Yamamoto}, \citenamefont {Haga}, \citenamefont {{\=O}nuki}, \citenamefont {Aoki}, \citenamefont {Matsuda}, \citenamefont {Sato},\ and\ \citenamefont {Takahashi}}]{hedo1998superconducting}%
  \BibitemOpen
  \bibfield  {author} {\bibinfo {author} {\bibfnamefont {M.}~\bibnamefont {Hedo}}, \bibinfo {author} {\bibfnamefont {Y.}~\bibnamefont {Inada}}, \bibinfo {author} {\bibfnamefont {E.}~\bibnamefont {Yamamoto}}, \bibinfo {author} {\bibfnamefont {Y.}~\bibnamefont {Haga}}, \bibinfo {author} {\bibfnamefont {Y.}~\bibnamefont {{\=O}nuki}}, \bibinfo {author} {\bibfnamefont {Y.}~\bibnamefont {Aoki}}, \bibinfo {author} {\bibfnamefont {T.~D.}\ \bibnamefont {Matsuda}}, \bibinfo {author} {\bibfnamefont {H.}~\bibnamefont {Sato}},\ and\ \bibinfo {author} {\bibfnamefont {S.}~\bibnamefont {Takahashi}},\ }\bibfield  {title} {\bibinfo {title} {Superconducting properties of {CeRu}$_2$},\ }\href {https://doi.org/10.1143/JPSJ.67.272} {\bibfield  {journal} {\bibinfo  {journal} {J. Phys. Soc. Jpn.}\ }\textbf {\bibinfo {volume} {67}},\ \bibinfo {pages} {272} (\bibinfo {year} {1998})}\BibitemShut {NoStop}%
\bibitem [{\citenamefont {Ashcroft}\ and\ \citenamefont {Mermin}(1976)}]{AshcroftMermin1976}%
  \BibitemOpen
  \bibfield  {author} {\bibinfo {author} {\bibfnamefont {N.~W.}\ \bibnamefont {Ashcroft}}\ and\ \bibinfo {author} {\bibfnamefont {N.~D.}\ \bibnamefont {Mermin}},\ }\href@noop {} {\emph {\bibinfo {title} {Solid State Physics}}}\ (\bibinfo  {publisher} {Saunders College Publishing},\ \bibinfo {address} {Philadelphia},\ \bibinfo {year} {1976})\BibitemShut {NoStop}%
\bibitem [{\citenamefont {Pritchard~Cairns}(2021)}]{Cairns2021}%
  \BibitemOpen
  \bibfield  {author} {\bibinfo {author} {\bibfnamefont {L.~S.}\ \bibnamefont {Pritchard~Cairns}},\ }\emph {\bibinfo {title} {Thermal {Hall} and heat capacity studies of topological magnetic insulators and superconductors}},\ \href {https://doi.org/10.7488/era/957} {Ph.D. thesis},\ \bibinfo  {school} {University of Edinburgh} (\bibinfo {year} {2021})\BibitemShut {NoStop}%
\bibitem [{\citenamefont {Werthamer}\ \emph {et~al.}(1966)\citenamefont {Werthamer}, \citenamefont {Helfand},\ and\ \citenamefont {Hohenberg}}]{Werthamer_Helfand_Hohenberg_1966}%
  \BibitemOpen
  \bibfield  {author} {\bibinfo {author} {\bibfnamefont {N.~R.}\ \bibnamefont {Werthamer}}, \bibinfo {author} {\bibfnamefont {E.}~\bibnamefont {Helfand}},\ and\ \bibinfo {author} {\bibfnamefont {P.~C.}\ \bibnamefont {Hohenberg}},\ }\bibfield  {title} {\bibinfo {title} {Temperature and purity dependence of the superconducting critical field, {$H_{c2}$}. {III}. {Electron} spin and spin-orbit effects},\ }\href {https://doi.org/10.1103/PhysRev.147.295} {\bibfield  {journal} {\bibinfo  {journal} {Phys. Rev.}\ }\textbf {\bibinfo {volume} {147}},\ \bibinfo {pages} {295} (\bibinfo {year} {1966})}\BibitemShut {NoStop}%
\bibitem [{\citenamefont {Caroli}\ and\ \citenamefont {Cyrot}(1965)}]{Caroli1965}%
  \BibitemOpen
  \bibfield  {author} {\bibinfo {author} {\bibfnamefont {C.}~\bibnamefont {Caroli}}\ and\ \bibinfo {author} {\bibfnamefont {M.}~\bibnamefont {Cyrot}},\ }\bibfield  {title} {\bibinfo {title} {Thermal conductivity in dirty superconducting alloys in high field},\ }\href {https://doi.org/10.1007/BF02422842} {\bibfield  {journal} {\bibinfo  {journal} {Phys. Kondens. Mater.}\ }\textbf {\bibinfo {volume} {4}},\ \bibinfo {pages} {285} (\bibinfo {year} {1965})}\BibitemShut {NoStop}%
\bibitem [{\citenamefont {Zhang}\ \emph {et~al.}(2026)\citenamefont {Zhang}, \citenamefont {Li}, \citenamefont {Behnia},\ and\ \citenamefont {Zhu}}]{zhang_2026_thermal_hall_niobium}%
  \BibitemOpen
  \bibfield  {author} {\bibinfo {author} {\bibfnamefont {J.}~\bibnamefont {Zhang}}, \bibinfo {author} {\bibfnamefont {X.}~\bibnamefont {Li}}, \bibinfo {author} {\bibfnamefont {K.}~\bibnamefont {Behnia}},\ and\ \bibinfo {author} {\bibfnamefont {Z.}~\bibnamefont {Zhu}},\ }\href {https://doi.org/10.48550/arXiv.2609.15663} {\bibinfo {title} {Thermal {Hall} effect in elemental niobium, a two-gap superconductor}} (\bibinfo {year} {2026}),\ \Eprint {https://arxiv.org/abs/2609.15663} {arXiv:2609.15663} \BibitemShut {NoStop}%
\bibitem [{\citenamefont {Wolf}\ \emph {et~al.}(1997)\citenamefont {Wolf}, \citenamefont {Hinkel}, \citenamefont {Holtmeier}, \citenamefont {Kouroudis}, \citenamefont {Bruls}, \citenamefont {Hiraoka}, \citenamefont {Assmus}, \citenamefont {Sun},\ and\ \citenamefont {L{\"u}thi}}]{Wolf1997}%
  \BibitemOpen
  \bibfield  {author} {\bibinfo {author} {\bibfnamefont {B.}~\bibnamefont {Wolf}}, \bibinfo {author} {\bibfnamefont {C.}~\bibnamefont {Hinkel}}, \bibinfo {author} {\bibfnamefont {S.}~\bibnamefont {Holtmeier}}, \bibinfo {author} {\bibfnamefont {I.}~\bibnamefont {Kouroudis}}, \bibinfo {author} {\bibfnamefont {G.}~\bibnamefont {Bruls}}, \bibinfo {author} {\bibfnamefont {T.}~\bibnamefont {Hiraoka}}, \bibinfo {author} {\bibfnamefont {W.}~\bibnamefont {Assmus}}, \bibinfo {author} {\bibfnamefont {W.}~\bibnamefont {Sun}},\ and\ \bibinfo {author} {\bibfnamefont {B.}~\bibnamefont {L{\"u}thi}},\ }\bibfield  {title} {\bibinfo {title} {Elastic properties of {CeRu}$_2$},\ }\href {https://doi.org/10.1007/BF02397466} {\bibfield  {journal} {\bibinfo  {journal} {J. Low Temp. Phys.}\ }\textbf {\bibinfo {volume} {107}},\ \bibinfo {pages} {421} (\bibinfo {year} {1997})}\BibitemShut {NoStop}%
\bibitem [{\citenamefont {Gross}\ \emph {et~al.}(1986)\citenamefont {Gross}, \citenamefont {Chandrasekhar}, \citenamefont {Einzel}, \citenamefont {Andres}, \citenamefont {Hirschfeld}, \citenamefont {Ott}, \citenamefont {Beuers}, \citenamefont {Fisk},\ and\ \citenamefont {Smith}}]{gross_1986_gap_interpolation}%
  \BibitemOpen
  \bibfield  {author} {\bibinfo {author} {\bibfnamefont {F.}~\bibnamefont {Gross}}, \bibinfo {author} {\bibfnamefont {B.~S.}\ \bibnamefont {Chandrasekhar}}, \bibinfo {author} {\bibfnamefont {D.}~\bibnamefont {Einzel}}, \bibinfo {author} {\bibfnamefont {K.}~\bibnamefont {Andres}}, \bibinfo {author} {\bibfnamefont {P.~J.}\ \bibnamefont {Hirschfeld}}, \bibinfo {author} {\bibfnamefont {H.~R.}\ \bibnamefont {Ott}}, \bibinfo {author} {\bibfnamefont {J.}~\bibnamefont {Beuers}}, \bibinfo {author} {\bibfnamefont {Z.}~\bibnamefont {Fisk}},\ and\ \bibinfo {author} {\bibfnamefont {J.~L.}\ \bibnamefont {Smith}},\ }\bibfield  {title} {\bibinfo {title} {Anomalous temperature dependence of the magnetic field penetration depth in superconducting {UBe}$_{13}$},\ }\href {https://doi.org/10.1007/BF01303700} {\bibfield  {journal} {\bibinfo  {journal} {Z. Phys. B}\ }\textbf {\bibinfo {volume} {64}},\ \bibinfo {pages} {175} (\bibinfo {year} {1986})}\BibitemShut {NoStop}%
\bibitem [{\citenamefont {Callaway}(1959)}]{callaway_1959_lattice_thermal_conductivity}%
  \BibitemOpen
  \bibfield  {author} {\bibinfo {author} {\bibfnamefont {J.}~\bibnamefont {Callaway}},\ }\bibfield  {title} {\bibinfo {title} {Model for lattice thermal conductivity at low temperatures},\ }\href {https://doi.org/10.1103/PhysRev.113.1046} {\bibfield  {journal} {\bibinfo  {journal} {Phys. Rev.}\ }\textbf {\bibinfo {volume} {113}},\ \bibinfo {pages} {1046} (\bibinfo {year} {1959})}\BibitemShut {NoStop}%
\bibitem [{\citenamefont {Berman}(1976)}]{Berman1976}%
  \BibitemOpen
  \bibfield  {author} {\bibinfo {author} {\bibfnamefont {R.}~\bibnamefont {Berman}},\ }\href@noop {} {\emph {\bibinfo {title} {Thermal Conduction in Solids}}},\ Oxford Studies in Physics\ (\bibinfo  {publisher} {Clarendon Press},\ \bibinfo {address} {Oxford},\ \bibinfo {year} {1976})\BibitemShut {NoStop}%
\bibitem [{\citenamefont {Lyu}\ \emph {et~al.}(2025)\citenamefont {Lyu}, \citenamefont {Cairns}, \citenamefont {Rodriguez}, \citenamefont {Liu}, \citenamefont {Ng}, \citenamefont {Singleton},\ and\ \citenamefont {Analytis}}]{Lyu2025}%
  \BibitemOpen
  \bibfield  {author} {\bibinfo {author} {\bibfnamefont {Y.}~\bibnamefont {Lyu}}, \bibinfo {author} {\bibfnamefont {L.~P.}\ \bibnamefont {Cairns}}, \bibinfo {author} {\bibfnamefont {J.}~\bibnamefont {Rodriguez}}, \bibinfo {author} {\bibfnamefont {C.}~\bibnamefont {Liu}}, \bibinfo {author} {\bibfnamefont {K.}~\bibnamefont {Ng}}, \bibinfo {author} {\bibfnamefont {J.}~\bibnamefont {Singleton}},\ and\ \bibinfo {author} {\bibfnamefont {J.~G.}\ \bibnamefont {Analytis}},\ }\bibfield  {title} {\bibinfo {title} {Entanglement randomness and gapped itinerant carriers in a frustrated quantum magnet},\ }\href {https://doi.org/10.1103/tx6t-gbxy} {\bibfield  {journal} {\bibinfo  {journal} {Phys. Rev. X}\ }\textbf {\bibinfo {volume} {15}},\ \bibinfo {pages} {041035} (\bibinfo {year} {2025})}\BibitemShut {NoStop}%
\bibitem [{\citenamefont {Tewordt}\ and\ \citenamefont {W{\"o}lkhausen}(1989)}]{tewordtTheoryThermalConductivity1989}%
  \BibitemOpen
  \bibfield  {author} {\bibinfo {author} {\bibfnamefont {L.}~\bibnamefont {Tewordt}}\ and\ \bibinfo {author} {\bibfnamefont {T.}~\bibnamefont {W{\"o}lkhausen}},\ }\bibfield  {title} {\bibinfo {title} {Theory of thermal conductivity of the lattice for high-{$T_c$} superconductors},\ }\href {https://doi.org/10.1016/0038-1098(89)90509-7} {\bibfield  {journal} {\bibinfo  {journal} {Solid State Commun.}\ }\textbf {\bibinfo {volume} {70}},\ \bibinfo {pages} {839} (\bibinfo {year} {1989})}\BibitemShut {NoStop}%
\end{thebibliography}
\end{document}